\documentclass[10pt]{article}

\usepackage{amsmath,amssymb,amsfonts,graphicx,amscd,amsfonts,epsfig,bm}
\usepackage{hyperref} 
\usepackage{cleveref} 
\usepackage{autonum} 
\usepackage{subfig}
\usepackage{cite}
\usepackage{enumitem}
\usepackage[svgnames]{xcolor}
\usepackage{hyperref}
\usepackage{float}
\usepackage{ulem}

\numberwithin{equation}{section}

\allowdisplaybreaks[4]

\begin{document}
\begin{titlepage}
\setcounter{page}{1001}
\renewcommand{\thefootnote}{\fnsymbol{footnote}}
\begin{normalsize}
\begin{flushright}
\begin{tabular}{l}
UTHEP-808\\
\end{tabular}
\end{flushright}
\end{normalsize}

~~\\

\vspace*{0cm}
    \begin{Large}
       \begin{center}
         {Derivation of the NS5-brane limit of the plane wave matrix model}
       \end{center}
    \end{Large}
\vspace{0.7cm}

\begin{center}
Yuhma A\textsc{sano}$^{1),2)}$\footnote[1]
            {
e-mail address : 
asano@het.ph.tsukuba.ac.jp},
Goro I\textsc{shiki}$^{1),2)}$\footnote[2]
            {
e-mail address : 
ishiki@het.ph.tsukuba.ac.jp}
and
Shinji S\textsc{himasaki}\footnote[4]
            {
e-mail address : 
shimasaki.s@gmail.com}

\vspace{0.7cm}

     $^{ 1)}$ {\it Graduate School of Science and Technology, University of Tsukuba, }\\
               {\it Tsukuba, Ibaraki 305-8571, Japan}\\

     $^{ 2)}$ {\it Tomonaga Center for the History of the Universe, University of Tsukuba, }\\
               {\it Tsukuba, Ibaraki 305-8571, Japan}\\
                                  
               \end{center}

\vspace{0.5cm}

\begin{abstract}
\noindent
From the gauge/gravity duality, it was predicted that there exists a nontrivial double scaling limit of 
the plane wave matrix model (the BMN matrix model),
which describes the type IIA little string theory (LST) on $R\times S^5$. 
We show on the gauge theory side that such a limit indeed exists 
for the partition function and in a certain 1/4 BPS sector of the matrix model,
and consequently derive an eigenvalue integral,
which is expected to describe the 1/4 BPS sector of LST.
\end{abstract}

\end{titlepage}

\tableofcontents

\section{Introduction}
String theory contains various extended objects \cite{Polchinski:1995mt}. 
Though understanding the whole dynamics of those objects would be 
an ultimate goal, this would need a non-perturbative formulation of
string theory, which has not been established yet.
As a first step towards this goal, understanding the decoupling limit, in which 
all the bulk interactions are 
turned off and there only remain interactions on each extended objects, should be important.
Such a limit is also very interesting in the context of the gauge/gravity 
correspondence \cite{Maldacena:1997re,Gubser:1998bc,Witten:1998qj, Itzhaki:1998dd}.

For D-branes, it is well-known that the dynamics in the decoupling limit 
is governed by the super Yang-Mills theory.
In contrast, the theory on NS5-branes, which is called the little string theory (LST) 
\cite{Aharony:1999ks,Kutasov:2001uf}, has been little understood. 
Assuming the gauge/gravity correspondence, one can see some properties of LST from the gravity description. 
However, any direct Lagrangian description of LST has not been established yet.

An interesting proposal for this problem was given in the context of 
the gauge/gravity correspondence for theories with $SU(2|4)$ symmetry \cite{Lin:2005nh}.
In this correspondence, the gauge theory side consists of a class of $SU(2|4)$ symmetric theories, 
which contains the type IIA LST on $R \times S^5$ and the plane wave matrix model (PWMM) also known as the BMN matrix model
\cite{Berenstein:2002jq}.
By investigating the dual gravitational backgrounds for these theories, 
it was found that their dual geometries are related via a sort of double scaling limit \cite{Ling:2006up}.
It was thus predicted that the corresponding double scaling limit exists on the gauge theory side in a nontrivial manner.
The significance of this prediction is that, if such a double scaling limit exists,
a Lagrangian formulation of LST on $R \times S^5$ would be given 
in terms of the double scaled PWMM with the well-defined Lagrangian formulation.

This prediction was made based on the gauge/gravity correspondence, which is still a conjecture. 
Hence, investigating whether the limit exists solely on the gauge theory side is important not only 
for establishing the Lagrangian formulation of LST but also providing nontrivial evidence 
of the gauge/gravity correspondence.
In \cite{Asano:2022brd}, the existence of the limit was numerically confirmed for 
the partition function and a 1/4 BPS sector of PWMM. 
In this paper, we prove the existence of the double scaling limit 
by performing an exact analysis for a 1/4 BPS sector of PWMM using the method developed in \cite{Volin:2009tqx,Marino:2019fuy,Reichert:2020ymc},
which enables us to derive the strong coupling expansions of relevant observables.
In the double scaling limit, we obtain an eigenvalue integral,
which is expected to describe the 1/4 BPS sector of LST.
The eigenvalue integral is interpreted as describing D0-branes in the NS5-brane background on the gravity side,
and the fixed quantity in the limit can be identified with an effective string coupling, which is the maximum value for the dilaton.
By using this eigenvalue integral, one can readily perform a perturbative analysis of LST in the weak coupling regime,
and even in the strong coupling regime some features of LST may remain tractable.

This paper is organized as follows. In section 2, we review some known aspects
of PWMM. 
In section 3, we analyze the partition function of PWMM
around a nontrivial background.
In section 4, we show the existence of the nontrivial double scaling limit of the partition
function and the 1/4 BPS sector.
In section 5, we summarize our result and discuss future directions.


\section{Some known aspects of the plane wave matrix model} \label{Some known aspects of the plane wave matrix model}
In this section, we review PWMM \cite{Berenstein:2002jq} 
and summarize some known aspects relevant to this paper.

The bosonic part of the action of PWMM in the Euclidean signature is given by
\begin{align}
S= \frac{1}{g^2} \int d\tau \int {\rm Tr}
\left( 
\frac{1}{2}(DX_M)^2 + \frac{1}{2}\left(\tilde{m} X_a -\frac{i}{2}\epsilon_{abc}[X_a, X_b] \right)^2
+ \frac{\tilde{m}^2}{8 }X_m^2
-\frac{1}{2}[X_a, X_m]^2 -\frac{1}{4}[X_m, X_n]^2
\right).
\end{align}
Here, $M =1,2, \cdots, 9$, $a, b, c=1,2,3$ and $m, n =4, 5, \cdots, 9$.
The one dimensional covariant derivative is defined as usual as 
$DX= \partial_\tau -i [A, X]$ with the gauge field $A$.
$\tilde{m}$ and $g$ are the mass parameter and the coupling constant of the model, 
respectively. $X_a$ are $\tau$-dependent Hermitian matrices and we 
denote the matrix size by $N$.
Since $\tilde{m}$ and $g^2$ has the mass dimension 1 and 3, respectively, 
only the dimensionless ratio $g^2/ \tilde{m}^3$ is relevant. 
We put $\tilde{m}=2$ in this paper to fix this redundancy.

Since the potential energy in the above action is written as a sum of positive 
definite terms, it is easy to find the classical vacuum.
It is given by  
\begin{align}
X_a = -2 L_a
\label{classical fuzzy sphere vacuum}
\end{align}
and all the other fields are vanishing, where $L_a$ are $N$-dimensional 
representation matrices of the $SU(2)$ generators, satisfying
$[L_a, L_b] = i\epsilon_{abc}L_c$. 
As long as the total dimension is $N$, any representation gives a classical vacuum. 
Thus, the vacuum is labeled by a partition of $N$. 
More specifically, with the irreducible decomposition, we can write $L_a$ as
\begin{align}
L_a = \bigoplus_{s=1}^{\nu } 1_{N_s} \otimes L_a^{[n_s]},
\label{irreducible decomposition}
\end{align}
where $n_s$ and $N_s$ are the dimension and the multiplicity 
of the $s$th irreducible representation and $\nu$ is the number of
kinds of irreducible representations contained in $L_a$.
Here, we assume $n_1 > n_2 > \cdots > n_\nu$ to avoid overcounting.
As long as $\sum_{s=1}^{\nu} N_s n_s =N$, 
(\ref{classical fuzzy sphere vacuum})
gives a classical vacuum for any $\nu$ and any
$\{(N_s, n_s)\}_{s=1, 2, \cdots, \nu }$.
Since the theory is massive, each classical vacuum is perturbatively stable 
and is labeled by the discrete moduli parameters, $\nu$ and $\{(N_s, n_s)\}_{s=1, 2, \cdots, \nu }$.
Furthermore, the classical vacua preserve all the 16 supersymmetries of 
the theory and hence are protected from quantum corrections.
Thus, (\ref{classical fuzzy sphere vacuum}) gives the quantum mechanical 
vacua of PWMM \cite{Dasgupta:2002hx,Kim:2002if,Dasgupta:2002ru}, 
and it makes sense to consider PWMM around every fixed vacuum.

For each of such theories, the gravity dual geometry was constructed in \cite{Lin:2005nh}.
It was shown on the gravity side that the BPS equation for solutions preserving $SU(2|4)$ symmetry is identical to 
the Laplace equation for a three-dimensional axially symmetric electrostatic system,
consisting of a background potential and charged conducting disks,
and finding the gravity solution reduces to solving for the electrostatic potential.
The gravity solution has D2- and NS5-brane charges, which correspond, respectively, 
to the charge and position of the conducting disks in the electrostatic system.
On the gauge theory side, they are related to the moduli parameters $\{(N_s, n_s)\}_{s=1,2\cdots,\nu}$ \cite{Maldacena:2002rb,Lin:2004nb,Lin:2005nh}.

Intriguingly, the dual geometry possesses a double scaling limit, in which
the geometry reduces to a spherical NS5-brane solution.
In \cite{Ling:2006up}, the form of the limit was identified 
for the trivial vacuum (with $L_a =0$) and rewritten in 
terms of the parameters on the gauge theory side. 
This work was further generalized in \cite{Asano:2022brd} for general vacua.
The form of the double scaling limit was then found to be
\begin{align}
N_1\to \infty, \; 
\lambda_1:=g^2N_1\to\infty, \quad \text{with}\quad N_1\lambda_1^{-\frac{5}{8}}e^{-\frac{\pi(8\lambda_1)^{\frac{1}{4}}}{n_1}}=\text{fixed},
\label{DSL intro}
\end{align}
and the other moduli parameters are kept fixed, where $\lambda_1$ is the 't Hooft coupling 
made from the multiplicity of the largest representation in (\ref{irreducible decomposition}).
Then, we see that the above limit is totally different from any known limit in large-$N$ gauge theories, such as the 't Hooft limit.
Hence, even apart from the consistency check of the gauge/gravity correspondence, 
it is interesting to show the existence of the above limit in the large-$N$ PWMM.

In \cite{Asano:2022brd}, the existence of the limit \eqref{DSL intro} was numerically confirmed for a certain 1/4-BPS sector of PWMM. 
In that work, by using the localization developed in \cite{Asano:2012zt}, 
the partition function of the matrix model was first reduced to a certain eigenvalue integral.
Then, by performing a Monte Carlo simulation of the eigenvalue integral, 
it was found that some 1/4-BPS operators have a consistent scaling law with the limit \eqref{DSL intro}.
In this paper, we start from the same eigenvalue integral and analytically show 
the existence of the double scaling limit.

For completeness, let us briefly review the eigenvalue integral, which describes the 1/4-BPS sector of PWMM for each fixed vacuum.
In the path-integral formulation, in order to define the theory around a fixed vacuum
\eqref{classical fuzzy sphere vacuum}, it is appropriate to adopt the boundary condition
such that all fields approach the vacuum configuration as $\tau \rightarrow \pm \infty$.
Under this boundary condition, by making a Wick rotation for the $X^9$-direction, 
one can construct 4 supersymmetries which leaves the following complex field invariant,
\begin{align}
\phi(\tau) = -X_4 (\tau) + X_8 (\tau) \sinh \tau +i X_9 (\tau) \cosh \tau.
\label{quarter_BPS_field}
\end{align}
By using these supersymmetries, one can perform the localization computation 
and reduce the partition function of PWMM to the following eigenvalue integral \cite{Asano:2012zt}: 
\begin{align}
&\int \prod_{s=1}^{\nu}\prod_{i=1}^{N_s}dq_{si} \times e^{-\frac{2}{g^2}\sum_s n_s\sum_i q_{si}^2} \nonumber\\
&\qquad \times \prod_{s,t=1}^{\nu}\prod_{J_{st}=\frac{| n_s-n_t |}{2}}^{\frac{n_s+n_t}{2}-1} \prod_{i_s=1}^{N_s}\prod_{j_t=1}^{N_t}
\hspace{-5.5mm} {\phantom{\prod}}^{\prime}
\left[\frac{\left\{(q_{si_s}-q_{tj_t})^2+(2J_{st})^2\right\}\left\{(q_{si_s}-q_{tj_t})^2+(2J_{st}+2)^2\right\}}
{\left\{(q_{si_s}-q_{tj_t})^2+(2J_{st}+1)^2\right\}^2}\right]^{\frac{1}{2}},
\label{starting eigenvalue integral}
\end{align}
where the remaining variables, $q_{si}$, are some moduli of the complex field $\phi$ arising in the localization. 
Here, $\prod'_{j_t}$ denotes the product that does not include $j_t=i_s$ if $t=s$.
From the general principle of the localization, any correlation functions of $\phi$ can also be computed with the above integral. 
It is noteworthy that the eigenvalue distributions of \eqref{starting eigenvalue integral} defined in the 't Hooft limit with the large $n_s$ and the large coupling 
can be mapped to the charge distributions of the conducting disks in the electrostatic problem associated to the dual gravity solution
\cite{Asano:2014vba, Asano:2014eca, Asano:2017xiy, Asano:2017nxw}.


\section{Plane wave matrix model around a vacuum with $\nu=1$}

In this section, we examine the eigenvalue integral for PWMM around a vacuum consisting of a single irreducible representation of $SU(2)$,
specifically, the case where $\nu=1$ in \eqref{starting eigenvalue integral}. 
We are interested in the parameter region where the 't Hooft coupling is large.
The eigenvalue integral \eqref{starting eigenvalue integral} in this region was studied in \cite{Asano:2014vba, Asano:2014eca, Asano:2017xiy, Asano:2017nxw},
where it was shown that the eigenvalue distribution satisfies an integral equation.
Here, we investigate the eigenvalue integral in more detail.
We first derive the loop equation at finite 't Hooft coupling and then solve for the eigenvalue distribution using a systematic expansion in the large coupling regime.
We also analytically examine the edge behavior of the distribution, 
which has not been extensively studied before but is crucial for determining the analytic form of the double scaling limit.
Throughout this section, for simplicity, we denote the dimension $n_1$ and its multiplicity $N_1$ of the irreducible representation of $SU(2)$
by $n$ and $N$, respectively.

\subsection{Loop equation}

The eigenvalue integral \eqref{starting eigenvalue integral}  for $\nu=1$ is given by
\begin{align}
Z=\int \prod_{i=1}^{N}dx_i \prod_{J=0}^{n-1} \prod_{i\neq j}
\left[\frac{\left\{(x_i-x_j)^2+(2J)^2\right\}\left\{(x_i-x_j)^2+(2J+2)^2\right\}}{\left\{(x_i-x_j)^2+(2J+1)^2\right\}^2}\right]^{\frac{1}{2}} 
e^{-\frac{2n}{g^2} \sum_i x_i^2}.
\label{Z nu=1}
\end{align}
The eigenvalue distribution and the resolvent are defined by
\begin{align}
&\rho(x)=\sum_{i}\delta(x-x_i) \quad (x\in \mathbb{R}), \\
&R(z)=\sum_i\frac{1}{z-x_i} = \int_{-\infty}^{\infty} dx \frac{\rho(x)}{z-x} \quad (z\in \mathbb{C}),
\end{align}
where $\int_{-\infty}^{\infty} dx \rho(x)=N$.
We examine \eqref{Z nu=1} in the large-$N$ limit with $\lambda=g^2N=\text{fixed}$.
The expectation values of the eigenvalue distribution and the resolvent in the large $N$ limit 
can be systematically obtained from the loop equation, which follows from the equality
\small
\begin{align}
\int \prod_{i=1}^{N}dx_i \sum_k \frac{\partial}{\partial x_k} \left( \frac{1}{z-x_k}
 \prod_{J=0}^{n-1} \prod_{i\neq j}
\left[\frac{\bigl\{(x_i-x_j)^2+(2J)^2\bigr\}\bigl\{(x_i-x_j)^2+(2J+2)^2\bigr\}}{\bigl\{(x_i-x_j)^2+(2J+1)^2\bigr\}^2}\right]^{\frac{1}{2}} 
e^{-\frac{2n}{g^2}\sum_i x_i^2}
\right)=0,
\end{align}
\normalsize
where $z\in\mathbb{C}$.
After some manipulations, we obtain the loop equation 
\begin{align}
\left\langle R(z)^2\right\rangle 
+\int dx \frac{1}{z-x}\left\langle \tilde{\bm{\mathsf{D}}}_nR(x) \rho(x)\right\rangle 
-\int dx \frac{nNV'(x)}{z-x}\langle \rho(x)\rangle =0, \label{Loop eq}
\end{align}
where $V(x)\equiv \frac{2}{\lambda}x^2$, $V^{\prime}(x)=\frac{d}{dx}V(x)=\frac{4}{\lambda}x$,
\small
\begin{align}
\tilde{\bm{\mathsf{D}}}_nR(x)
&\equiv R(x+2i)-2R(x+i) +\sum_{J=1}^{n-1}\Big[R(x+2Ji)+R(x+(2J+2)i)-2R(x+(2J+1)i)\Big] +\mathrm{c.c.},
\label{def tildeDR}
\end{align}
\normalsize
and $\langle\cdots\rangle$ stands for the expectation value with respect to \eqref{Z nu=1}.
Using the shift operator $\mathsf{D}\equiv e^{i\partial_x}$, which acts as $\mathsf{D}R(x)=R(x+i)$,
we can rewrite the right-hand side of \eqref{def tildeDR} in a more compact form as
\begin{align}
\tilde{\bm{\mathsf{D}}}_nR(x) 
&= \biggl[\mathsf{D}^{2}- 2\mathsf{D}+(1-\mathsf{D})^2\sum_{J=1}^{n-1}\mathsf{D}^{2J}\biggr]R(x+i\epsilon)+\mathrm{c.c.},
\end{align}
where $\epsilon$ is a positive infinitesimal quantity taken to zero.
By taking the difference between \eqref{Loop eq} with $z=x+i\epsilon$ and that with $z=x-i\epsilon$ ($x\in \mathbb{R}$), 
we obtain 
\begin{align}
 \left\langle \hat{\bm{\mathsf{D}}}_nR(x) \rho(x)\right\rangle - nNV'(x)\langle \rho(x)\rangle =0, \label{Loop eq 2}
\end{align}
where
\begin{align}
\hat{\bm{\mathsf{D}}}_nR(x) 
&\equiv \sum_{J=0}^{n-1}\Big[R(x+2Ji)+R(x+(2J+2)i)-2R(x+(2J+1)i)\Big] +\mathrm{c.c.} \nonumber\\
&= (1-\mathsf{D})^2\sum_{J=0}^{n-1}\mathsf{D}^{2J} R(x+i\epsilon)+\mathrm{c.c.}  \nonumber\\
&\equiv \frac{1-\mathsf{D}}{1+\mathsf{D}}(1-\mathsf{D}^{2n})R(x+i\epsilon)+\mathrm{c.c.} \ .
\label{Dhat}
\end{align}
The final expression in \eqref{Dhat} should be understood as being defined by the middle expression. 
Similar expressions appearing in the following sections are understood in the same manner.
We denote the large $N$ expansion of the eigenvalue distribution and the resolvent by
\begin{align}
&\langle \rho(x)\rangle=\sum_{g=0}^{\infty}\frac{1}{N^{2g-1}}\rho_{g}(x), \\
&\langle R(z)\rangle=\sum_{g=0}^{\infty}\frac{1}{N^{2g-1}}R_{g}(z), 
\end{align}
Note that $\langle\rho(x)\rangle=\langle\rho(-x)\rangle$ because of the symmetry of the integrand in \eqref{Z nu=1} under the simultaneous sign flip of $x_i$.
Hereafter, we restrict ourselves to the one-cut solution, assuming that $\langle\rho(x)\rangle$ takes values only within 
 a single support, $x\in [-x_m, x_m]$.

\subsection{Solution at large $N$}

In order to derive the equation for $R_0(z)$, we examine the leading contribution in the large $N$ expansion of the left-hand side of \eqref{Loop eq 2}.
The leading contribution is of $\mathcal{O}(N^2)$ and is given by
\begin{align}
\left(\hat{\bm{\mathsf{D}}}_nR_0(x)-nV'(x)\right)\rho_0(x)=0,  \label{Loop eq 3}
\end{align}
When $x\in [-x_m,x_m]$, since $\rho_0(x)\neq 0$, \eqref{Loop eq 3} then becomes
\begin{align}
 \hat{\bm{\mathsf{D}}}_nR_0(x)-nV'(x)&=0.
 \end{align}
 Using \eqref{Dhat} and $V(x)=\frac{2}{\lambda}x^2$, this can be written explicitly as
\begin{align}
\frac{1-\mathsf{D}}{1+\mathsf{D}}(1-\mathsf{D}^{2n}) R_0(x+i0)+ \frac{1-\mathsf{D}^{- 1}}{1+\mathsf{D}^{- 1}}(1-\mathsf{D}^{- 2n})R_0(x-i0) &= \frac{4n}{\lambda}x. 
\label{Loop eq leading}
\end{align}

Since our interest is to show the double scaling limit \eqref{DSL intro}, we consider the parameter regime where the 't Hooft coupling is large, $\lambda\gg 1$,
and consequently, the extent of the eigenvalues is also large, $x_m\gg 1$.
To solve \eqref{Loop eq leading} in this regime, we follow the method developed in \cite{Volin:2009tqx,Marino:2019fuy,Reichert:2020ymc}, 
in which perturbative expansion in $x_m\to \infty$ is analyzed in two regimes:
\begin{align}
&\text{bulk regime:}\qquad x_m\to \infty \quad \text{and}\quad x\to \infty \quad \text{with} \quad \frac{x}{x_m}=\text{fixed}, \\
&\text{edge regime:}\qquad x_m\to \infty \quad \text{and}\quad x\to \infty \quad \text{with} \quad \frac{x-x_m}{n}=\text{fixed}.
\end{align}
This approach enables a systematic expansion at large $\lambda$.
Here, we present a brief summary of the results, whereas a detailed analysis is provided in Appendix \ref{app:detail_solution}.
In the large $\lambda$ limit, $x_m$ is given by
\begin{align}
x_m&=\left(8\lambda\right)^\frac{1}{4}. \label{xm solution large N}
\end{align}
The regular part of $R_{0}(z)$, which has no discontinuity and satisfies \eqref{Loop eq leading}, is determined as
\begin{align}
R_{0,\mathrm{reg}}(z)=\frac{4}{x_m^2}z-\frac{8}{3x_m^4}z^3.
\end{align}
While $R_{0,\mathrm{reg}}(x)$ is common in both regimes, 
the discontinuous part, $R_{0, \mathrm{disc}}(z)\equiv R_{0}(z)-R_{0,\mathrm{reg}}(z)$ depends on the bulk or edge regime.
In the bulk regime, $R_{0,\mathrm{disc}}(z)$ is given by
\begin{align}
R_{0,\mathrm{disc}}(z)&=\frac{8}{3x_m^4}(z^2-x_m^2)^{\frac{3}{2}}. \label{R0 bulk sol}
\end{align}
In the edge regime, we focus on $R_{0,\mathrm{disc}}(x_m+nt)$,
where 
$t$ represents the coordinate around the edge $t\equiv \frac{x-x_m}{n}$.
We denote the Laplace transform of $R_{0,\mathrm{disc}}(x_m+nt)$ with respect to $t$ by $\hat{R}_{0}(s)$, 
\begin{align}
R_{0,\mathrm{disc}}(x_m+nt)=\int_0^{\infty} ds\, e^{-ts} \hat{R}_{0}(s). \label{laplace transf R0 edge}
\end{align}
It is shown in Appendix \ref{app:solution of the full equation} that $\hat{R}_{0}(s)$ is of $\mathcal{O}(x_m^{-\frac{5}{2}})$ and takes the form
\begin{align}
\hat{R}_{0}(s)
&=
\frac{4\sqrt{2}n^{\frac{3}{2}}}{x_m^{\frac{5}{2}}} 
 \frac{1}{s^{\frac{5}{2}}}
 \exp\left[\frac{s}{\pi}\log\left(\frac{\pi e}{s}\right)\right]
\Gamma\left(\frac{s}{\pi}+1\right)
\frac{\Gamma\left(\frac{s}{2n\pi}+1\right)}{\Gamma\left(\frac{s}{2n\pi}+\frac{1}{2}\right)},
\label{R0(s) edge sol}
\end{align}
In the next section, we will use this result to determine the form of the double scaling limit.

Before closing this section, we note the connection to the large $N$ analysis of \eqref{Z nu=1} performed in \cite{Asano:2014vba, Asano:2014eca, Asano:2017xiy, Asano:2017nxw}.
While the large $N$ loop equation \eqref{Loop eq leading} is valid for finite $n$ and finite $\lambda$,
the corresponding equation obtained in \cite{Asano:2014vba, Asano:2014eca, Asano:2017xiy, Asano:2017nxw} assumes the limit
\begin{align}
n\to \infty, \;\; x_m\to\infty \; (\lambda\to\infty) \;\; \text{with} \;\; \frac{n}{x_m}=\text{fixed},
\end{align}
and therefore takes a different form from \eqref{Loop eq leading}.
In the context of the gravity dual, this limit corresponds to taking the D0-brane charge large 
and taking a typical scale of the geometry larger than the string scale, so that the classical geometry description becomes reliable.
One can verify that their equation can be obtained from \eqref{Loop eq leading} by taking the same limit, in which the following approximation becomes valid,
\begin{align}
\frac{1-\mathsf{D}^{\pm1}}{1+\mathsf{D}^{\pm1}}
=\mp i \left(\frac{1}{2}\partial_x + \frac{1}{24}\partial_x^3+\cdots\right).
\label{(1-D)/(1+D) expansion}
\end{align}
Under this approximation, \eqref{Loop eq leading} can be written as
\begin{align}
&(1-\mathsf{D}^{2n}) R_0(x+i\epsilon)-(1-\mathsf{D}^{- 2n})R_0(x-i\epsilon) = -2i\mu+i\frac{4n}{\lambda}x^2, 
\label{Loop eq leading approx}
\end{align}
where $\mu$ is an integration constant.
In terms of the eigenvalue distribution, \eqref{Loop eq leading approx} can be also rewritten as\footnote{
This is identical to the integral equation that governs the one-hole excitation energy 
of the Lieb-Liniger model \cite{Lieb:1963rt,Lieb:1963zz,Korepin:1993kvr}, and also appears in other contexts, for example \cite{Zarembo:2008hb}.
The integral equation obtained by replacing the right-hand side of \eqref{int eq rho0} with a constant
is known as the Love integral equation \cite{Love1949THEEF,Hutson1963TheCP} or the Lieb-Liniger integral equation \cite{Lieb:1963rt}.}
\begin{align}
\rho_0(x)-\frac{1}{\pi}\int_{-x_m}^{x_m} dy\frac{2n}{(x-y)^2+(2n)^2}\rho_0(y)=\frac{\mu}{\pi}-\frac{2n}{\pi\lambda}x^2.
\label{int eq rho0}
\end{align}
The eigenvalue extent $x_m$ and the integration constant $\mu$ are determined from $\int_{-x_m}^{x_m} dx\,\rho_0(x)=1$ and $\rho_0(x_m)=0$, 
as shown in Appendix \ref{solution of fredholm eq}. 
This equation is exactly the same as the one derived in \cite{Asano:2014vba, Asano:2014eca, Asano:2017xiy, Asano:2017nxw},
and is shown to be equivalent to that for the charge distribution function characterizing the dual gravity solution.
Thus, \eqref{Loop eq leading} may be regarded as their finite $n$ generalization.

\section{Double scaling limit in the plane wave matrix model}
In this section, we show the existence of the double scaling limit \eqref{DSL intro}
in \eqref{starting eigenvalue integral} for $\nu\geq 2$.
We will demonstrate that, in the limit, the eigenvalues $q_{1i}\: (i=1,\cdots,N_1)$ can be integrated out,
yielding an effective theory for the remaining eigenvalues $q_{si_s} \: (s\geq 2, i_s=1,\cdots,N_s)$. 
We will then show that the saddle point equation of the effective theory coincides with
the equation determining the gravity dual of LST on $R\times S^5$ derived in \cite{Asano:2022brd}. 
Hereafter, we denote the eigenvalue distributions and the resolvents by
\begin{align}
\rho^{(s)}(x)&\equiv \sum_{i=1}^{N_s}\delta(x-q_{si}) \quad (x\in \mathbb{R}), \\
\quad R^{(s)}(z)&\equiv \sum_{i=1}^{N_s}\frac{1}{z-q_{si}}=\int dx \frac{\rho^{(s)}(x)}{z-x} \quad (z \in \mathbb{C}). \label{dist q}
\end{align}


\subsection{Double scaling limit for a vacuum with $\nu=2$} \label{sec:DSL_nu=2}

First, let us consider the eigenvalue integral for PWMM around a vacuum with $\nu=2$. 
In this case, the eigenvalue integral is given by
\begin{align}
Z&=\int \prod_{i=1}^{N_1} dx_i \prod_{a=1}^{N_2} dy_a  \; e^{-\frac{2n_1}{g^2}\sum_ix_{i}^2-\frac{2n_2}{g^2}\sum_ay_a^2} \nonumber\\
&\qquad \times \prod_{J=0}^{n_1-1}\prod_{i\neq j}
\left[\frac{\left\{(x_i-x_j)^2+(2J)^2\right\}\left\{(x_i-x_j)^2+(2J+2)^2\right\}}{\left\{(x_i-x_j)^2+(2J+1)^2\right\}^2}\right]^{\frac{1}{2}} \nonumber\\
&\qquad\times \prod_{J=0}^{n_2-1}\prod_{a\neq b}
\left[\frac{\left\{(y_a-y_b)^2+(2J)^2\right\}\left\{(y_a-y_b)^2+(2J+2)^2\right\}}{\left\{(y_a-y_b)^2+(2J+1)^2\right\}^2}\right]^{\frac{1}{2}} \nonumber\\
&\qquad\times \prod_{J=\frac{|n_1-n_2|}{2}}^{\frac{n_1+n_2}{2}-1}\prod_{i}
\prod_{a}\frac{\left\{(x_i-y_a)^2+(2J)^2\right\}\left\{(x_i-y_a)^2+(2J+2)^2\right\}}{\left\{(x_i-y_a)^2+(2J+1)^2\right\}^2}.
\label{Z_2disk}
\end{align}
We denote hermitian matrices with eigenvalues $x_i \: (i=1,\cdots,N_1)$ and $y_a \: (a=1,\cdots,N_2)$ by $X$ and $Y$, respectively.
In the double scaling limit, $N_1$ and $\lambda_1=g^2N_1$ are taken to be large.
In the dual gravity description, this corresponds to a configuration characterized by  D0-branes residing in an NS5-brane throat,
where the matrix $X$ represents the NS5-brane background
and the matrix $Y$ the D0-branes \cite{Lin:2005nh,Asano:2014vba,Asano:2014eca}\footnote{
Note that the situation here is quite different from the situation studied in \cite{Asano:2017xiy,Asano:2017nxw}, 
where both $N_1$ and $N_2$ are taken to be large with $g^2N_1$ and $g^2N_2$ held fixed.
In this case, there are no degrees of freedom trapped within the throat region.
}. 
In the gauge theory description, the matrix $Y$ may be interpreted as a low-energy degree of freedom of LST on $R\times S^5$.

In order to verify that the double scaling limit indeed exists in the strong coupling region, $\lambda_1\gg 1$,
we focus solely on the integral over $x_i$ and define
\begin{align}
Z(Y)&\equiv \int \prod_{i=1}^{N_1} dx_i  \; e^{-\frac{2N_1n_1}{\lambda_1}\sum_i x_{i}^2}
 \prod_{J=0}^{n_1-1}\prod_{i\neq j}
\left[\frac{\left\{(x_i-x_j)^2+(2J)^2\right\}\left\{(x_i-x_j)^2+(2J+2)^2\right\}}{\left\{(x_i-x_j)^2+(2J+1)^2\right\}^2}\right]^{\frac{1}{2}}  \nonumber\\
&\qquad \times  \prod_{J=\frac{|n_1-n_2|}{2}}^{\frac{n_1+n_2}{2}-1}
\prod_{i}\prod_{a}\frac{\left\{(x_i-y_a)^2+(2J)^2\right\}\left\{(x_i-y_a)^2+(2J+2)^2\right\}}{\left\{(x_i-y_a)^2+(2J+1)^2\right\}^2}.
\label{Z(Y)}
\end{align}
Here $y_a$ are treated as external variables. 
Note that, in terms of the free energy $F(Y)=\log Z(Y)$, the full eigenvalue integral \eqref{Z_2disk} can be written as
\begin{align}
Z
&=\int \prod_{a=1}^{N_2} dy_a \prod_{J=0}^{n_2-1}\prod_{a\neq b}
\left[\frac{\left\{(y_a-y_b)^2+(2J)^2\right\}\left\{(y_a-y_b)^2+(2J+2)^2\right\}}{\left\{(y_a-y_b)^2+(2J+1)^2\right\}^2}\right]^{\frac{1}{2}}
e^{-\frac{2n_2}{g^2}\sum_{a}y_a^2+F(Y)}.
\label{Z_2disk_2}
\end{align}
Later, we will demonstrate that the eigenvalues $y_a$ around the origin are subject to an attractive potential of the form $\cosh(\frac{\pi y}{n_1})$, 
indicating that $y=0$ is a critical point of the potential.
In this paper, we examine the double scaling limit in the case where $y_a$ are distributed 
sufficiently close to this critical point by comparison with the extent of $x_i$.

\subsubsection{Large $N_1$ expansion}

Let us examine the large $N_1$ expansion of the free energy  $F(Y)=\log Z(Y)$.
We view $Z(Y)$ as the expectation value of a determinant operator. 
Specifically, $Z(Y)$ can be written as
\begin{align}
Z(Y)=Z_X\left\langle \mathrm{det}\: C(X,Y)\right\rangle_X,
\label{Z(Y)_2}
\end{align}
where $\langle\cdots\rangle_X$ denotes the expectation value with respect to $Z_X$ 
defined by
\begin{align}
Z_X=\int \prod_{i=1}^{N_1} dx_i  \; e^{-N_1n_1\sum_iV(x_{i})}
 \prod_{J=0}^{n_1-1}\prod_{i\neq j}
\left[\frac{\left\{(x_i-x_j)^2+(2J)^2\right\}\left\{(x_i-x_j)^2+(2J+2)^2\right\}}{\left\{(x_i-x_j)^2+(2J+1)^2\right\}^2}\right]^{\frac{1}{2}},
\end{align}
with $V(x)=\frac{2}{\lambda_1}x^2$,
and
\begin{align}
\mathrm{det}\: C(X,Y)\equiv \prod_{i}\prod_{a}\prod_{J=\frac{|n_1-n_2|}{2}}^{\frac{n_1+n_2}{2}-1}
\frac{\left\{(x_i-y_a)^2+(2J)^2\right\}\left\{(x_i-y_a)^2+(2J+2)^2\right\}}{\left\{(x_i-y_a)^2+(2J+1)^2\right\}^2}.
\label{detC}
\end{align}
Here, $\det C(X,Y)$ is interpreted as the interaction terms between $x_i$ and the external variables $y_a$.
Importantly, the insertion of $\det C(X,Y)$ introduces holes in two-dimensional surfaces that correspond to 
dual Feynman diagrams for $Z_X$ \cite{Ginsparg:1993is,Marino:2012zq,Ishiki:2010wb}.
This can be easily seen from the following expression of the free energy obtained by 
 taking the logarithm of \eqref{Z(Y)_2},
\begin{align}
F(Y)=\log Z_X+\sum_{h=1}^{\infty}\frac{1}{h!} \left\langle \big(\mathrm{Tr} \log C(X,Y)\big)^h\right\rangle_{X,c},
\end{align}
where we have used $\det C(X,Y)=e^{\mathrm{Tr}\log C(X,Y)}$ 
and $\langle \cdots \rangle_{X,c}$ is the connected part of $\langle \cdots \rangle_{X}$.

As a result, the large $N_1$ expansion of the free energy $F(Y)$ is given by 
\begin{align}
F(Y)\equiv \sum_{g=0}^{\infty}\sum_{h=0}^{\infty} N_1^{2-2g-h}F_{gh}(Y), \label{F large N}
\end{align}
where $g$ and $h$ represent the numbers of the genus and the holes, respectively\footnote{
Note that the argument so far is analogous to that for eigenvalue instantons in matrix models 
\cite{David:1992za,Shenker:1990uf,Hanada:2004im,Marino:2012zq}.
For example, in a one-matrix model with multiple critical points, 
an isolated eigenvalue in a one-cut background is identified as an instanton.
The eigenvalues $y_a$ in \eqref{Z(Y)_2} are reminiscent of such isolated eigenvalues.
However, the key difference is that $y_a$ are not separated eigenvalues from $x_i$ 
even though both originate from the same moduli field \eqref{quarter_BPS_field}.
It is not clear whether $y_a$ can be interpreted as the eigenvalue instantons.}. 
Specifically, 
$F_{gh}(Y) \; (h\geq 1)$ is given by the genus $g$ component of $\langle \big(\mathrm{Tr} \log C(X,Y)\big)^h\rangle_{X,c}$,
while $F_{g0}$  corresponds to the genus $g$ component of $F_X\equiv \log Z_X$,
with the expansion,
\begin{align}
F_X= \log Z_X=\sum_{g=0}^{\infty} N_1^{2-2g}F_{g0}. \label{F_X}
\end{align}
One can obtain the correlation functions $\langle \big(\mathrm{Tr} \log C(X,Y)\big)^h\rangle_{X,c}$ from those of the resolvent $R^{(1)}(z)$.
The correlation function $\langle R^{(1)}(z_1)\cdots R^{(1)}(z_h)\rangle_{X,c}$
can, in principle, be computed by applying the iterative procedure based on 
the large $N_1$ expansion of the loop equation 
and the loop insertion operator \cite{Ambjorn:1992jf,Ambjorn:1992gw}, although the explicit calculation seems to be difficult.

To calculate the free energy $F(Y)$, let us adopt a different perspective.
It is clear from \eqref{Z(Y)} that the derivative of $F(Y)$ with respect to $y_a$, a particular eigenvalue of $Y$, is given by 
\begin{align}
\frac{\partial}{\partial y_a}F(Y)
=\left\langle \hat{\bm{\mathsf{D}}}_{n_1,n_2}{R}^{(1)}(y_a)
\right\rangle ,
\label{del_yF=barR}
\end{align}
where $\langle\cdots\rangle$ stands for the expectation value with respect to $Z(Y)$ and
$\hat{\bm{\mathsf{D}}}_{n_1,n_2}{R}^{(1)}(z)$ is defined by 
\begin{align}
\hat{\bm{\mathsf{D}}}_{n,m}R(x)
&\equiv \sum_{J=\frac{|n-m|}{2}}^{\frac{n+m}{2}-1}\Big[R(x+2Ji)+R(x+(2J+2)i)-2R(x+(2J+1)i)\Big]+\mathrm{c.c.} \nonumber\\
&= (1-\mathsf{D})^2\sum_{J=\frac{|n-m|}{2}}^{\frac{n+m}{2}-1}\mathsf{D}^{2J}R(x+i\epsilon)+\mathrm{c.c.} \nonumber\\
&\equiv \frac{1-\mathsf{D}}{1+\mathsf{D}}(\mathsf{D}^{|n-m|}-\mathsf{D}^{n+m})R(x+i\epsilon)+\mathrm{c.c.} .
\label{Dhat2}
\end{align}
Note that \eqref{Dhat2} with $m=n$ reduces to \eqref{Dhat}.
One can expand $\langle R^{(1)}(z)\rangle$ as
\begin{align}
\left\langle R^{(1)}(z)\right\rangle 
&\equiv \sum_{g=0}^{\infty}\sum_{h=0}^{\infty} N_1^{1-2g-h} R^{(1)}_{gh}(z).
\label{R large N}
\end{align}
Then, the following relation holds for $h\geq 1$,
\begin{align}
&\frac{\partial}{\partial y_a}F_{gh}(Y)
=\hat{\bm{\mathsf{D}}}_{n_1,n_2}{R}_{gh-1}^{(1)}(y_a).
\label{F_gh and R_gh}
\end{align}
Therefore, the free energy is obtained by calculating the expectation values of $\hat{\bm{\mathsf{D}}}_{n_1,n_2}{R}^{(1)}(y)$,
which can be derived using the loop equation.

\subsubsection{Loop equation}

Let us examine the loop equation for \eqref{Z(Y)}, which is given by
\begin{align}
&\left\langle R^{(1)}(z)^2\right\rangle 
+\int dx \frac{1}{z-x}\left\langle \tilde{\bm{\mathsf{D}}}_{n_1}R^{(1)}(x) \rho^{(1)}(x)
\right\rangle  \nonumber\\
&\quad +\int dx \frac{1}{z-x} \hat{\bm{\mathsf{D}}}_{n_1,n_2}R^{(2)}(x)\left\langle \rho^{(1)}(x)\right\rangle
 -n_1N_1\int dx \frac{V'(x)}{z-x}\langle \rho^{(1)}(x)\rangle =0.
\label{Loop eq 2disk}
\end{align}
Taking the difference in \eqref{Loop eq 2disk} between $z=x+i\epsilon$ and $z=x-i\epsilon$ ($x\in \mathbb{R}$), we obtain
\begin{align}
&\left\langle \hat{\bm{\mathsf{D}}}_{n_1}R^{(1)}(x) \rho^{(1)}(x)\right\rangle  
+\hat{\bm{\mathsf{D}}}_{n_1,n_2}R^{(2)}(x) \left\langle \rho^{(1)}(x)\right\rangle
-n_1N_1V'(x)\langle \rho^{(1)}(x)\rangle =0.
\label{Loop eq 2disk simple}
\end{align}
We then perform the large $N_1$ expansion for the correlation functions of the resolvent $R^{(1)}(z)$ and $\rho^{(1)}(x)$,
and keep the terms up to $\mathcal{O}(N_1^0)$, where the connected part of the correlation functions does not appear, 
as\footnote{
The large $N_1$ expansion for $\rho^{(1)}(x)$ is given in a similar manner to that for $R^{(1)}(x)$ in \eqref{R large N}.
}
\begin{align}
&N_1^2\left(\hat{\bm{\mathsf{D}}}_{n_1} R^{(1)}_{00}(x)-n_1V'(x)\right)\rho^{(1)}_{00}(x) \nonumber\\
&\; +N_1\left\{\left(\hat{\bm{\mathsf{D}}}_{n_1} R^{(1)}_{00}(x)-n_1V'(x)\right)\rho^{(1)}_{01}(x)
+\left(\hat{\bm{\mathsf{D}}}_{n_1}R^{(1)}_{01}(x)+\hat{\bm{\mathsf{D}}}_{n_1,n_2}R^{(2)}(x)\right)\rho^{(1)}_{00}(x)\right\}
\nonumber\\
&\; +\mathcal{O}(N_1^0) =0.
\end{align}

\subsubsection*{$\bullet$ loop equation at $\mathcal{O}(N_1^2)$}

The leading term of the loop equation in the large $N_1$ expansion is of $\mathcal{O}(N_1^2)$ and takes the following form,
\begin{align}
(\hat{\bm{\mathsf{D}}}_{n_1}R^{(1)}_{00}(x)-n_1V'(x))\rho^{(1)}_{00}(x)=0.
\label{loop eq 1}
\end{align}
This is the same equation as \eqref{Loop eq leading} examined in the previous section.
As before, we suppose that $\rho^{(1)}_{00}(x)$ is nonzero for $x\in [-x_m, x_m]$ and zero otherwise.
\eqref{loop eq 1} then reduces to
\begin{align}
\hat{\bm{\mathsf{D}}}_{n_1} R^{(1)}_{00}(x)-\frac{4n_1}{\lambda_1}x=0, \qquad x\in [-x_m, x_m], \label{loop eq 1-2}
\end{align}
which can be solved in the same way as described below \eqref{Loop eq leading}.

\subsubsection*{$\bullet$ loop equation at $\mathcal{O}(N_1)$}

The next-to-leading term of the loop equation in the large $N_1$ expansion is of $\mathcal{O}(N_1)$ and takes the following form,
\begin{align}
\left(\hat{\bm{\mathsf{D}}}_{n_1} R^{(1)}_{00}(x)-n_1V'(x)\right)\rho^{(1)}_{01}(x)
+\left(\hat{\bm{\mathsf{D}}}_{n_1}R^{(1)}_{01}(x)+\hat{\bm{\mathsf{D}}}_{n_1,n_2}R^{(2)}(x)\right)\rho^{(1)}_{00}(x)=0.
\label{loop eq 2}
\end{align}
By using \eqref{loop eq 1-2}, 
it reduces to
\begin{align}
\hat{\bm{\mathsf{D}}}_{n_1} R^{(1)}_{01}(x)+\hat{\bm{\mathsf{D}}}_{n_1,n_2}R^{(2)}(x)=0, \qquad x\in [-x_m, x_m] .
\label{loop eq 2-2}
\end{align}

\subsubsection{Calculation of $F_{01}(Y)$}

As shown in \eqref{F_gh and R_gh}, $F_{01}(Y)$ can be derived from
\begin{align}
\frac{\partial}{\partial y_a}F_{01}(Y)=\hat{\bm{\mathsf{D}}}_{n_1,n_2}R^{(1)}_{00}(y_a). \label{F01_R00}
\end{align}
We evaluate the right-hand side by using \eqref{loop eq 1-2} and the following formula, which is valid for $n_1>n_2$,\footnote{
For $n_1<n_2$, $S_{n_1,n_2}(x)$ takes the different form \eqref{S n1<n2},
and the following argument no longer holds.
}
\begin{align}
&\hat{\bm{\mathsf{D}}}_{n_1,n_2}R^{(1)}(y)=\int_{-\infty}^{\infty} dx\, S_{n_1,n_2}(y-x)\hat{\bm{\mathsf{D}}}_{n_1}R^{(1)}(x),
\label{hatR1 to hatR2}
\end{align}
where
\begin{align}
S_{n_1,n_2}(y-x)
&\equiv \frac{1}{2\pi i}\sum_{k=-\infty}^{\infty}\left(\frac{1}{y+i(n_1+n_2)-x+2in_1k}-\frac{1}{y+i(n_1-n_2)-x+2in_1k}\right) \label{Sdef}\\
&= \frac{1}{2n_1}\frac{\sin\frac{\pi n_2}{n_1}}{\cosh\frac{\pi(y-x)}{n_1}+\cos\frac{\pi n_2}{n_1}} \\
&=\frac{1}{n_1}\sum_{k=1}^{\infty}(-1)^{k-1}e^{-\frac{k\pi |y-x|}{n_1}}\sin\frac{k\pi n_2}{n_1}.
\label{S}
\end{align}
The derivation of this formula can be found in Appendix \ref{app:derivation_S}.
Using 
\begin{align}
\hat{\bm{\mathsf{D}}}_{n_1} R^{(1)}_{00}(x)=
\begin{cases}
n_1V'(x) \quad &(|x|\leq x_m), \\
n_1V'(x)+\hat{\bm{\mathsf{D}}}_{n_1}R^{(1)}_{00,\mathrm{disc}}(x) \quad &(|x|\geq x_m),
\end{cases}
\end{align}
the right-hand side of \eqref{F01_R00} is written as
\begin{align}
\hat{\bm{\mathsf{D}}}_{n_1,n_2}R^{(1)}_{00}(y)
&=\int_{-\infty}^{\infty} dx\, S_{n_1,n_2}(y-x)\hat{\bm{\mathsf{D}}}_{n_1}R^{(1)}_{00}(x) \nonumber\\
&= \int_{-\infty}^{\infty} dx\, S_{n_1,n_2}(y-x)\: n_1V'(x) 
+ \int_{x_m}^{\infty} dx (S_{n_1,n_2}(y-x)-S_{n_1,n_2}(y+x))\hat{\bm{\mathsf{D}}}_{n_1}R^{(1)}_{00,\mathrm{disc}}(x), 
\label{hatD2 R}
\end{align}
where $x_m=(8\lambda)^{\frac{1}{4}}$, as in \eqref{xm solution large N}.
The first term can be evaluated as
\begin{align}
\text{the 1st term of \eqref{hatD2 R}}
= \int_{-\infty}^{\infty} dx\, S_{n_1,n_2}(y-x)n_1V'(x) 
= \frac{4n_2}{\lambda_1}y.
\end{align}
The second term is rewritten as
\begin{align}
\text{the 2nd term of \eqref{hatD2 R}}
&=\int_{x_m}^{\infty} dx (S_{n_1,n_2}(y-x)-S_{n_1,n_2}(y+x))\hat{\bm{\mathsf{D}}}_{n_1}R^{(1)}_{00,\mathrm{disc}}(x) \nonumber\\
&=\frac{2}{n_1}\int_{x_m}^{\infty} dx\, e^{-\frac{\pi x}{n_1}}\hat{\bm{\mathsf{D}}}_{n_1}R^{(1)}_{00,\mathrm{disc}}(x)
\times \sinh{\frac{\pi y}{n_1}}\sin\frac{\pi n_2}{n_1} 
+ \mathcal{O}\left(e^{-\frac{2\pi x_m}{n_1}}\right),
\label{2nd term of barR1}
\end{align}
where in the second line we have used \eqref{S} and kept only the term with $k=1$ 
based on the assumption that the extent of $y_a$ is much smaller than $x_m$, as noted below \eqref{Z_2disk_2}.
Note that, because of the exponential suppression factor, the integral receives dominant contributions from values of $x$ near $x_m$.
Therefore, it can be evaluated by using the edge-regime approximation of the discontinuous part $R^{(1)}_{00,\mathrm{disc}}(x)$, 
as described in \eqref{laplace transf R0 edge} and \eqref{R0(s) edge sol}.
We then obtain
\begin{align}
\frac{2}{n_1}\int_{x_m}^{\infty} dx\, e^{-\frac{\pi x}{n_1}}\hat{\bm{\mathsf{D}}}_{n_1}R^{(1)}_{00,\mathrm{disc}}(x) 
&=2e^{-\frac{\pi x_m}{n_1}} \int_{0}^{\infty} dt\, e^{-\pi t}\hat{\bm{\mathsf{D}}}_{n_1}R^{(1)}_{00,\mathrm{disc}}(x)|_{x=x_m+n_1t} \nonumber\\
&=-\frac{A(n_1)}{x_m^{\frac{5}{2}}}e^{-\frac{\pi x_m}{n_1}}+\mathcal{O}\left(e^{-\frac{2\pi x_m}{n_1}}\right) ,
\end{align}
where $A(n_1)$ is defined by
\begin{align}
A(n_1)\equiv 16\sqrt{2}n_1^{\frac{3}{2}}
\int_0^{\infty}ds
\frac{\tan\frac{s}{2n_1}\sin 2s}{s+\pi}\frac{1}{s^{\frac{5}{2}}}e^{\frac{s}{\pi}\log\frac{\pi e}{s}}\Gamma\left(\frac{s}{\pi}+1\right)
\frac{\Gamma\left(\frac{s}{2n_1\pi}+1\right)}{\Gamma\left(\frac{s}{2n_1\pi}+\frac{1}{2}\right)}.
\label{A}
\end{align}
Note that for $n_1\gg 1$, $A(n_1)$ is proportional to $\sqrt{n_1}$.
Using \eqref{F01_R00}
together with \eqref{hatD2 R}--\eqref{A} and the fact that $F(Y)$ is totally symmetric in $\{y_a\}$,  
we obtain $F_{01}(Y)$ as
\begin{align}
F_{01}(Y)=\frac{2n_2}{\lambda_1}\sum_{a=1}^{N_2}y_a^2
-\frac{n_1 A(n_1) e^{-\frac{\pi x_m}{n_1}}}{\pi x_m^{\frac{5}{2}}} \sin\frac{\pi n_2}{n_1}  \sum_{a=1}^{N_2}\cosh{\frac{\pi y_a}{n_1}} 
+ \mathcal{O}\left(e^{-\frac{2\pi x_m}{n_1}}\right),
\label{F01 result}
\end{align}
where we have ignored a $y_a$-independent constant term.

\subsubsection{Calculation of $F_{02}(Y)$}

As shown in \eqref{F_gh and R_gh}, $F_{02}(Y)$ can be derived from
\begin{align}
\frac{\partial}{\partial y_a}F_{02}(Y)=\hat{\bm{\mathsf{D}}}_{n_1,n_2} R^{(1)}_{01}(y_a). \label{F02_R01}
\end{align}
As in the previous subsection, we evaluate the right-hand side by using \eqref{loop eq 2-2} and \eqref{hatR1 to hatR2} as
\begin{align}
\hat{\bm{\mathsf{D}}}_{n_1,n_2}R^{(1)}_{01}(y)
&=\int_{-\infty}^{\infty} dx\, S_{n_1,n_2}(y-x)\hat{\bm{\mathsf{D}}}_{n_1} R^{(1)}_{01}(x) \nonumber\\
&= -\int_{-\infty}^{\infty} dx\, S_{n_1,n_2}(y-x)\:  \hat{\bm{\mathsf{D}}}_{n_1,n_2} R^{(2)}(x) \nonumber\\
&\qquad + \int_{x_m}^{\infty} dx (S_{n_1,n_2}(y-x)-S_{n_1,n_2}(y+x))\left(
\hat{\bm{\mathsf{D}}}_{n_1} R^{(1)}_{01}(x)+\hat{\bm{\mathsf{D}}}_{n_1,n_2} R^{(2)}(x)\right).
\label{evaluate barR_01}
\end{align}
By using \eqref{Sdef}, the first term can be evaluated as
\begin{align}
&-\int_{-\infty}^{\infty} dx\, S_{n_1,n_2}(y-x)\:  \hat{\bm{\mathsf{D}}}_{n_1,n_2}R^{(2)}(x) \nonumber\\
&=-\frac{1}{2\pi i}\int_{-\infty}^{\infty} dx \sum_{k=-\infty}^{\infty}\left(\frac{1}{y+i((2k+1)n_1+n_2)-x}-\frac{1}{y+i((2k+1)n_1-n_2)-x}\right) 
\hat{\bm{\mathsf{D}}}_{n_1,n_2}R^{(2)}(x) \nonumber\\
&=\hat{\hat{\bm{\mathsf{D}}}}_{n_1,n_2}R^{(2)}(y),
\label{-S hatD2 R}
\end{align}
where
\begin{align}
\hat{\hat{\bm{\mathsf{D}}}}_{n,m}R(y)
&\equiv \sum_{k=-\infty, k\neq 0}^{\infty}\frac{1-\mathsf{D}}{1+\mathsf{D}}(\mathsf{D}^{|2kn|}-\mathsf{D}^{|2kn+2m|})R(y+i\epsilon)
+\mathrm{c.c.} .
\label{hathatD R}
\end{align}
Then, \eqref{-S hatD2 R} can be rewritten as
\small
\begin{align}
\hat{\hat{\bm{\mathsf{D}}}}_{n_1,n_2}R^{(2)}(y)
&=\frac{\partial}{\partial y}
\sum_{k=1}^{\infty}\left(
 \sum_{J=kn_1}^{kn_1+n_2-1} - \sum_{J=kn_1-n_2}^{kn_1-1}
\right)
\sum_{b=1}^{N_2}
\log \frac{\bigl\{(2J)^2+(y-y_b)^2\bigr\}\bigl\{(2J+2)^2+(y-y_b)^2\bigr\}}{\bigl\{(2J+1)^2+(y-y_b)^2\bigr\}^2}.
\label{hathatD R explicit 2}
\end{align}
\normalsize

In the evaluation of the second term of \eqref{evaluate barR_01}, 
we note the assumption that the extent of the eigenvalues $\{y_a\}$ is much smaller than $x_m$.
We also note that $R^{(1)}_{gh}(z)$ for general $g$ and $h$ should not increase exponentially in $x_m(=(8\lambda)^{1/4})$ as $\lambda$ becomes large.
This is because, through the iterative procedure \cite{Ambjorn:1992jf,Ambjorn:1992gw}, it is related to $R^{(1)}_{00}(z)$, 
which does not exhibit such exponential growth.
It then follows that the second term in \eqref{evaluate barR_01} is of $\mathcal{O}(e^{-\frac{\pi x_m}{n_1}})$.

Thus, using the symmetry of $F(Y)$, we obtain $F_{02}(Y)$ from \eqref{F02_R01}--\eqref{hathatD R explicit 2} as
\begin{align}
F_{02}(Y)= \frac{1}{2} \sum_{a \neq b} \sum_{k=1}^{\infty} 
\left(\sum_{J=kn_1}^{kn_1+n_2-1} - \sum_{J=kn_1-n_2}^{kn_1-1}\right)
\log \frac{\bigl\{(2J)^2+(y_a-y_b)^2\bigr\}\bigl\{(2J+2)^2+(y_a-y_b)^2\bigr\}}{\bigl\{(2J+1)^2+(y_a-y_b)^2\bigr\}^2}
+\mathcal{O}\left(e^{-\frac{\pi x_m}{n_1}}\right),
\label{F02 result}
\end{align}
where again, we have ignored a $y_a$-independent term.

\subsubsection{Double scaling limit}

Since $F_{gh}(Y)$ is expressed in terms of  the correlation functions of $R^{(1)}(z)$ with respect to $Z_X$ as noted below \eqref{F_X},
it should not exhibit exponential growth with respect to $x_m$ for the same reason as $R^{(1)}_{gh}(z)$.
Therefore, the contributions 
from $F_{gh}(Y)$ 
are of $\mathcal{O}(N_1^{-1})$
except for $F_{01}(Y)$ and $F_{02}(Y)$.
Consequently, substituting \eqref{F large N}, \eqref{F_X}, \eqref{F01 result} and \eqref{F02 result} into the exponent of the integrand in \eqref{Z_2disk_2},
we obtain
\small
\begin{align}
-\frac{2N_1n_2}{\lambda_1}\sum_{a=1}^{N_2}y_a^2+F(Y)
&=
F_{X}
-\frac{n_1 A(n_1) N_1e^{-\frac{\pi x_m}{n_1}}}{\pi x_m^{\frac{5}{2}}} \sin\frac{\pi n_2}{n_1} \sum_{a=1}^{N_2}\cosh{\frac{\pi y_a}{n_1}}
\nonumber\\
&\quad +\frac{1}{2} \sum_{a \neq b} \sum_{k=1}^{\infty} 
\left(\sum_{J=kn_1}^{kn_1+n_2-1} - \sum_{J=kn_1-n_2}^{kn_1-1}\right)
\log \frac{\bigl\{(2J)^2+(y_a-y_b)^2\bigr\}\bigl\{(2J+2)^2+(y_a-y_b)^2\bigr\}}{\bigl\{(2J+1)^2+(y_a-y_b)^2\bigr\}^2}
\nonumber\\
&\quad +\mathcal{O}\left(N_1e^{-\frac{2\pi x_m}{n_1}}, e^{-\frac{\pi x_m}{n_1}}, N_1^{-1}\right).
\label{exponent}
\end{align}
\normalsize
Note that the contribution from the first term of \eqref{F01 result} cancels the original gaussian potential and leaves the hyperbolic cosine potential,
as remarked below \eqref{Z_2disk_2}.
From this expression, we find a nontrivial double scaling limit,
\begin{align}
N_1\to\infty, \; \lambda_1\to\infty, \; n_1=\text{fixed}
\quad \text{with}  \quad
\frac{n_1 A(n_1) N_1e^{-\frac{\pi  (8\lambda_1)^{\frac{1}{4}}}{n_1}}}{\pi (8\lambda_1)^{\frac{5}{8}}}\equiv \frac{4}{\pi g_0}:\text{fixed},
\label{DSL}
\end{align}
in which \eqref{exponent} becomes
\small
\begin{align}
-\frac{2N_1n_2}{\lambda_1}\sum_{a=1}^{N_2}y_a^2+F(Y)  
&\to F_{X}-\frac{4}{\pi g_0} \sin\frac{\pi n_2}{n_1} \sum_{a=1}^{N_2}\cosh{\frac{\pi y_a}{n_1}} \nonumber\\
&\quad +\frac{1}{2} \sum_{a \neq b} \sum_{k=1}^{\infty} 
\left(\sum_{J=kn_1}^{kn_1+n_2-1} - \sum_{J=kn_1-n_2}^{kn_1-1}\right)
\log \frac{\bigl\{(2J)^2+(y_a-y_b)^2\bigr\}\bigl\{(2J+2)^2+(y_a-y_b)^2\bigr\}}{\bigl\{(2J+1)^2+(y_a-y_b)^2\bigr\}^2}.
\label{exponent in DSL}
\end{align}
\normalsize
As shown in section \ref{subsubsec: rho in DSL}, 
the fixed quantity $g_0$ is related to an effective string coupling, which is the maximum value of the dilaton, of the NS5-brane solution in the dual gravity \cite{Lin:2005nh}.
The constant factor $\frac{4}{\pi}$ on the right-hand side of \eqref{DSL} is introduced for convenience, to simplify the correspondence.

We focus on the $y_a$-dependent part in \eqref{exponent in DSL}.
Note that $F_X=\log Z_X$ \eqref{F_X} is divergent as $\mathcal{O}(N_1^2)$, but is independent of $y_a$. 
Hence, we consider \eqref{exponent in DSL} with $F_X$ subtracted, or equivalently, \eqref{Z_2disk_2} divided by $Z_X$. 
In conclusion, we find that in the double scaling limit \eqref{DSL}, the eigenvalue integral \eqref{Z_2disk_2} divided by $Z_X$ results in,
\begin{align}
Z_{\mathrm{DSL}}
&=\int \prod_ady_a \prod_{a\neq b} \left[ \prod_{J=0}^{n_2-1} \mathcal{B}(J,y_a-y_b)
\prod_{k=1}^{\infty}\left[
\prod_{J=kn_1}^{kn_1+n_2-1}\mathcal{B}(J,y_a-y_b)
\prod_{J=kn_1-n_2}^{kn_1-1}\mathcal{B}(J,y_a-y_b)^{-1}
\right] \right] \nonumber\\
&\qquad \times \exp\left({-\frac{4}{\pi g_0}\sin\frac{\pi n_2}{n_1}\sum_{a=1}^{N_2}\cosh\frac{\pi y_a}{n_1}}\right),
\label{Z_ns5limit}
\end{align}
up to an overall constant factor,
where 
\begin{align}
\mathcal{B}(J,y)\equiv \left[\frac{\bigl\{(2J)^2+y^2\bigr\}\bigl\{(2J+2)^2+y^2\bigr\}}{\bigl\{(2J+1)^2+y^2\bigr\}^2}\right]^{\frac{1}{2}}.
\end{align}
This eigenvalue integral is considered to describe the 1/4-BPS sector of LST on $R\times S^5$,
characterized by discrete parameters $n_1,n_2(<n_1), N_2$.
From the gravity perspective, $n_1$ corresponds to the number of NS5-branes forming the background geometry, 
while $N_2$ and $n_2$ represent D2- and NS5-brane fluxes in the background with the D0-brane charge given by $N_2n_2$ \cite{Lin:2005nh}. 
The correspondence becomes apparent in section \ref{subsubsec: rho in DSL}, where we examine the loop equation of \eqref{Z_ns5limit} 
in the large $N_2$ and large $n_2$ limits, and interpret it in the gravity dual.

While evaluating the eigenvalue integral \eqref{Z_ns5limit} seems difficult in general, there may be certain limits where it becomes more tractable.
\begin{itemize}
\item In the case of $g_0\ll 1$,
the expectation value of a BPS operator (such as $\operatorname{Tr}Y^k$) can be
computed perturbatively and obtained
as a power series in $g_0$, i.e.~$\sum_n a_ng_0^n$. 
This implies that in the large $N_1$ expansion of the original eigenvalue integral \eqref{Z_2disk}, 
the coefficient of $N_1^{-n}$ behaves as $(\lambda_1^{5/8} e^{\pi (8\lambda_1)^{1/4}/n_1})^n$ for large $\lambda_1$.
This behavior was originally conjectured in \cite{Ling:2006up} and later partially confirmed through numerical simulations \cite{Asano:2022brd}.
We have now rigorously proven it for the 1/4-BPS operators.

Note that the perturbative series in $g_0$ is calculable using a gaussian matrix model,
as the integrals over $y_a$ contribute mainly near $y_a\sim 0$.  
In fact, $Z_{\mathrm{DSL}}$ can be approximated at the leading order by 
\begin{align}
\int \prod_a dy_a \prod_{a>b}|y_a-y_b|^2 
\exp\left( -\frac{2\pi}{n_1^2 g_0}\sin\frac{\pi n_2}{n_1} \sum_{a=1}^{N_2} y_a^2 \right).
\end{align}
Here, we have neglected the constant term $-\frac{4N_2}{\pi g_0}\sin\frac{\pi n_2}{n_1}$ in the exponent,
because it does not affect on the calculation of expectation values.
However, this term has a possible physical interpretation, 
which we provide in section \ref{summary}.

\item  It is also interesting to consider a limit analogous to the 't Hooft limit: $N_2\to \infty, \:g_0\to 0 \; \mathrm{with}\; g_0N_2=\mathrm{fixed}$,
where the eigenvalues $y_a$ are expected to spread over a finite range.
In the gravity description, this corresponds to a situation in which the throat region contains nontrivial three- and six-cycles 
threaded by the NS5-brane and D2-brane fluxes, respectively.
In the gauge theory description, this is interpreted as a nontrivial vacuum of LST on $R\times S^5$ \cite{Lin:2005nh,Asano:2014vba,Asano:2014eca}.
In section \ref{subsubsec: rho in DSL},
we derive the loop equation for the eigenvalue distribution of $y_a$ in this limit
and show that it coincides with the equation that determines the gravity dual of LST.

\end{itemize}

\subsubsection{Eigenvalue distribution $\rho^{(2)}(y)$ in the double scaling limit} \label{subsubsec: rho in DSL}

Let us examine the eigenvalue distribution of $y_a$ in \eqref{Z_ns5limit}.
In the following, we denote the expectation value with respect to $Z_{\mathrm{DSL}}$ \eqref{Z_ns5limit} by $\langle \cdots \rangle_{\mathrm{DSL}}$.

The loop equation for $R^{(2)}(z)$ is given by
\begin{align}
&\left\langle R^{(2)}(z)^2\right\rangle_{\mathrm{DSL}}
+ \int dy'\frac{1}{z-y'}\left\langle \tilde{\bm{\mathsf{D}}}_{n_2} R^{(2)}(y')\rho^{(2)}(y')\right\rangle_{\mathrm{DSL}}
+ \int dy'\frac{1}{z-y'}\left\langle \hat{\hat{\bm{\mathsf{D}}}}_{n_1,n_2} R^{(2)}(y')\rho^{(2)}(y')\right\rangle_{\mathrm{DSL}} \nonumber\\
& \quad -\frac{4}{n_1 g_0}\sin\frac{\pi n_2}{n_1}\int dy\frac{\sinh\frac{\pi y'}{n_1}}{z-y'}\left\langle \rho^{(2)}(y')\right\rangle_{\mathrm{DSL}} =0,
\label{ns5 limit loop eq}
\end{align}
where $\hat{\hat{\bm{\mathsf{D}}}}_{n_1,n_2} R^{(2)}(y)$ is defined in \eqref{hathatD R}.
The difference in \eqref{ns5 limit loop eq} between $z=y+i\epsilon$ and $z=y-i\epsilon$ ($y\in \mathbb{R}$) is
\begin{align}
\left\langle \hat{\bm{\mathsf{D}}}_{n_2} R^{(2)}(y)\rho^{(2)}(y)\right\rangle_{\mathrm{DSL}}
+ \left\langle \hat{\hat{\bm{\mathsf{D}}}}_{n_1,n_2}R^{(2)}(y)\rho^{(2)}(y)\right\rangle_{\mathrm{DSL}}
-\frac{4}{n_1 g_0}\sin\frac{\pi n_2}{n_1}\sinh\frac{\pi y}{n_1}\left\langle\rho^{(2)}(y)\right\rangle_{\mathrm{DSL}} =0.
\label{ns5 limit loop eq 2}
\end{align}
We consider the large $N_2$ limit analogous to the 't Hooft limit, which is given by
\begin{align}
N_2\to \infty \quad \text{with}\quad t_2\equiv g_0N_2=\text{fixed},
\end{align} 
and expand $\left\langle R^{(2)}(z)\right\rangle_{\mathrm{DSL}}$ and $\left\langle \rho^{(2)}(y)\right\rangle_{\mathrm{DSL}}$ as
\begin{align}
\left\langle R^{(2)}(z)\right\rangle_{\mathrm{DSL}} &=\sum_{g=0}^{\infty}N_{2}^{1-2g}R^{(2)}_{\mathrm{DSL},g}(z), \nonumber\\
\left\langle \rho^{(2)}(y)\right\rangle_{\mathrm{DSL}} &=\sum_{g=0}^{\infty}N_{2}^{1-2g}\rho^{(2)}_{\mathrm{DSL},g}(y). \label{rho2 DSL expansion}
\end{align}
The leading contribution in the large $N_2$ expansion of \eqref{ns5 limit loop eq 2} is given by
\begin{align}
\left(\hat{\bm{\mathsf{D}}}_{n_2} R^{(2)}_{\mathrm{DSL},0}(y) 
+ \hat{\hat{\bm{\mathsf{D}}}}_{n_1,n_2} R_{\mathrm{DSL},0}^{(2)}(y)
-\frac{4}{n_1 t_2}\sin\frac{\pi n_2}{n_1}\sinh\frac{\pi y}{n_1}\right)\rho^{(2)}_{\mathrm{DSL},0}(y) =0.
\label{ns5 limit loop eq 3}
\end{align}
Assuming that $\rho_{\mathrm{DSL},0}^{(2)}(y)$ has nonzero support on $y\in [-y_m,y_m]$, 
then Eq.~\eqref{ns5 limit loop eq 3} reduces, within this interval, to
\begin{align}
&\sum_{k=-\infty}^{\infty} \left[\frac{1-\mathsf{D}}{1+\mathsf{D}} \left(\mathsf{D}^{|2kn_1|} - \mathsf{D}^{|2kn_1+2n_2|} \right)R^{(2)}_{\mathrm{DSL},0}(y+i\epsilon)
+\frac{1-\mathsf{D}^{-1}}{1+\mathsf{D}^{-1}}\left(\mathsf{D}^{-|2kn_1|} - \mathsf{D}^{-|2kn_1+2n_2|} \right)R^{(2)}_{\mathrm{DSL},0}(y-i\epsilon)\right] \nonumber\\
&\quad -\frac{4}{n_1 t_2}\sin\frac{\pi n_2}{n_1}\sinh\frac{\pi y}{n_1}=0.
\label{ns5 limit loop eq 4}
\end{align}
This is the large $N_2$ saddle point equation of \eqref{Z_ns5limit}.

Now, we consider the limit\footnote{
This does not contradict the double scaling limit \eqref{DSL}, which involves a nontrivial dependence on $n_1$. 
Indeed, for $n_1\gg 1$, the function $A(n_1)$ appearing in \eqref{DSL} and defined in \eqref{A} scales as $\sqrt{n_1}$.
Therefore, the relevant scaling quantities in \eqref{DSL} are $N_1/n_1$ and ${\lambda_1}^{\frac{1}{4}}/n_1$, 
and \eqref{DSL} can thus be regarded as the double scaling limit with respect to these quantities.}
\begin{align}
n_1\to \infty, \;\; n_2\to \infty, \;\; y_m\to\infty \;\; (t_2\to \infty) \;\; \text{with} \;\; \frac{n_1}{y_m}  \;\; \text{and} \;\;  \frac{n_2}{y_m} \;\; \text{held fixed},
\end{align}
which corresponds, in the gravity dual, to the regime in which the NS5-brane charges are large 
and a typical scale of the geometry is larger than the string scale, so that the classical geometry description becomes valid.
In this limit, $\frac{1-\mathsf{D}^{\pm 1}}{1+\mathsf{D}^{\pm 1}}$ can be expanded as in \eqref{(1-D)/(1+D) expansion},
and then we can rewrite \eqref{ns5 limit loop eq 4} in the form
\begin{align}
&\sum_{k=-\infty}^{\infty}\left[\left(\mathsf{D}^{|2kn_1|} - \mathsf{D}^{|2kn_1+2n_2|} \right)R^{(2)}_{\mathrm{DSL},0}(y+i\epsilon)
-\left(\mathsf{D}^{-|2kn_1|} - \mathsf{D}^{-|2kn_1+2n_2|} \right)R^{(2)}_{\mathrm{DSL},0}(y-i\epsilon)\right] \nonumber\\
&\quad = -2i\mu_2+i\frac{8}{\pi t_2}\sin\frac{\pi n_2}{n_1}\cosh\frac{\pi y}{n_1},
\label{DSL R2 eq}
\end{align}
or, in terms of the eigenvalue distribution $\rho^{(2)}_{\mathrm{DSL},0}(y)$, as 
\begin{align}
&\rho^{(2)}_{\mathrm{DSL},0}(y)
- \sum_{k=-\infty}^{\infty} \frac{1}{\pi} \int_{-y_m}^{y_m} dy' 
\left(
\frac{|2kn_1+2n_2|}{(y-y')^2+|2kn_1+2n_2|^2} - \frac{|2kn_1|}{(y-y')^2+|2kn_1|^2}\right)\rho^{(2)}_{\mathrm{DSL},0}(y') \nonumber\\
&\quad =\frac{\mu_2}{\pi} - \frac{4}{\pi^2 t_2}\sin\frac{\pi n_2}{n_1}\cosh\frac{\pi y}{n_1}.
\label{DSL rho2 eq}
\end{align}
where $\mu_2$ is an integration constant determined along with $y_m$ by $\int_{-y_m}^{y_m}dy\,\rho_{\mathrm{DSL},0}^{(2)}(y)=1$ 
and $\rho_{\mathrm{DSL},0}^{(2)}(\pm y_m)=0$. 
This is exactly the same integral equation that determines the gravity dual of a nontrivial vacuum of LST, 
where $\rho^{(2)}_{\mathrm{DSL},0}(y)$ corresponds to the charge distribution of a conducting disk in the associated electrostatic problem\footnote{
Interestingly, the integral equation \eqref{DSL rho2 eq} is identical to the thermodynamic Bethe ansatz equation 
for the principal chiral field model, which has been extensively studied in \cite{DiPietro:2021yxb}.
It would be intriguing to explore their results in our context.}.
It is then found that the fixed constant $g_0$ in \eqref{DSL} corresponds to the same parameter $g_0$ of the NS5-brane solution in \cite{Lin:2005nh,Asano:2022brd}
and is related to the maximum value of the dilaton (effective string coupling) as $g_s\sim g_0n_1^{\frac{3}{2}}$.

\subsection{Double scaling limit for general vacua}

The argument presented in the previous subsection can be readily extended to general vacua of PWMM.
As in the previous section, we rewrite the eigenvalue integral \eqref{starting eigenvalue integral} as
\begin{align}
Z&=\int  \prod_{s=2}^{\nu}\prod_{i=1}^{N_s}dq_{si} 
\times e^{-\sum_{s=2}^{\nu} \frac{2n_s}{g^2}\sum_{i=1}^{N_s} q_{si}^2} \nonumber\\
&\qquad \times \prod_{s,t=2}^{\nu}\prod_{J_{st}=\frac{| n_s-n_t |}{2}}^{\frac{n_s+n_t}{2}-1} \prod_{i_s=1}^{N_s}\prod_{j_t=1}^{N_t}
\hspace{-5.5mm} {\phantom{\prod}}^{\prime}
\left[\frac{\left\{(q_{si_s}-q_{tj_t})^2+(2J_{st})^2\right\}\left\{(q_{si_s}-q_{tj_t})^2+(2J_{st}+2)^2\right\}}
{\left\{(q_{si_s}-q_{tj_t})^2+(2J_{st}+1)^2\right\}^2}\right]^{\frac{1}{2}} Z(\{q\}), 
\label{Z general}
\end{align}
where $Z(\{q\})\equiv Z(q_2,q_3,\cdots,q_\nu)$ is the eigenvalue integral over $x_{i}\equiv q_{1i}$ defined by
\begin{align}
Z(\{q\})&\equiv \int \prod_{i=1}^{N_1} dx_i \; e^{-\frac{2n_1}{g^2}\sum_{i=1}^{N_1}x_i^2} 
 \prod_{J=0}^{n_1-1}\prod_{\substack{i, j=1\\ i\neq j}}^{N_1}
\left[\frac{\left\{(x_i-x_j)^2+(2J)^2\right\}\left\{(x_i-x_j)^2+(2J+2)^2\right\}}{\left\{(x_i-x_j)^2+(2J+1)^2\right\}^2}\right]^{\frac{1}{2}}  \nonumber\\
&\qquad \times \prod_{t=2}^{\nu}\prod_{J_{t}=\frac{| n_1-n_t |}{2}}^{\frac{n_1+n_t}{2}-1} \prod_{i=1}^{N_1}\prod_{j_t=1}^{N_t}
\left[\frac{\left\{(x_i-q_{tj_t})^2+(2J_{t})^2\right\}\left\{(x_i-q_{tj_t})^2+(2J_{t}+2)^2\right\}}{\left\{(x_i-q_{tj_t})^2+(2J_{t}+1)^2\right\}^2}\right].
\label{Z_qs}
\end{align}
In what follows, we will investigate the double scaling limit of the free energy $F(\{q\})=\log Z(\{q\})$.

The free energy can be calculated straightforwardly in parallel with subsection \ref{sec:DSL_nu=2} 
by making the replacements $F(Y)\to F(\{q\})$, $R^{(2)}(z)\to \sum_{s=2}^{\nu}R^{(s)}(z)$, and so on.
We assume that 
the free energy $F(\{q\})$ and the resolvent $R^{(1)}(z)$
have large $N_1$ expansions analogous to those in \eqref{F large N} and \eqref{R large N}.
The free energy at $\mathcal{O}(N_1)$  is given by
\begin{align}
F_{01}(\{q\})=\sum_{s=2}^{\nu}\sum_{i_s=1}^{N_s}\frac{2n_s}{\lambda_1}q_{si_s}^2
-\frac{n_1 A(n_1) e^{-\frac{\pi x_m}{n_1}}}{\pi x_m^{\frac{5}{2}}} \sum_{s=2}^{\nu}\sum_{i_s=1}^{N_s} \sin\frac{\pi n_s}{n_1} \cosh{\frac{\pi q_{si}}{n_1}} 
+ \mathcal{O}\left(e^{-\frac{2\pi x_m}{n_1}}\right) .
\label{general F01 result}
\end{align}
For the calculation of the free energy at $\mathcal{O}(N_1^0)$, we use, instead of \eqref{evaluate barR_01}--\eqref{hathatD R},
\begin{align}
\hat{\bm{\mathsf{D}}}_{n_1,n_s} R^{(1)}_{01}(q_{si})
&=\int_{-\infty}^{\infty} dx\, S_{n_1,n_s}(q_{si}-x)\hat{\bm{\mathsf{D}}}_{n_1} R^{(1)}_{01}(x) \nonumber\\
&=\sum_{t=2}^{\nu}\hat{\hat{\bm{\mathsf{D}}}}_{n_1,n_s,n_t}R^{(t)}(q_{si})+\mathcal{O}\left(e^{-\frac{\pi x_m}{n_1}}\right),
\label{evaluate barR_01 general}
\end{align}
where
\begin{align}
\hat{\hat{\bm{\mathsf{D}}}}_{n,m,l}R(x)
&\equiv 
\sum_{k=-\infty, k\neq 0}^{\infty} \frac{1-\mathsf{D}}{1+\mathsf{D}}(\mathsf{D}^{|2kn+m-l|} - \mathsf{D}^{|2kn+m+l|} )R(x+i\epsilon)
+\mathrm{c.c.} 
\label{hathatD R 2}
\end{align}
When $l=m$, Eq.~\eqref{hathatD R 2} reduces to \eqref{hathatD R}: $\hat{\hat{\bm{\mathsf{D}}}}_{n,m,m}=\hat{\hat{\bm{\mathsf{D}}}}_{n,m}$.
Note that the last expression in \eqref{evaluate barR_01 general} can be rewritten as 
\begin{align}
\hat{\hat{\bm{\mathsf{D}}}}_{n_1,n_s,n_t}R^{(t)}(y)
&=\frac{\partial}{\partial y} 
\sum_{k=1}^{\infty}\left(
 \sum_{J=kn_1+\frac{n_s-n_t}{2}}^{kn_1+\frac{n_s+n_t}{2}-1}
 -\sum_{J=kn_1-\frac{n_s+n_t}{2}}^{kn_1-\frac{n_s-n_t}{2}-1}
\right) 
\sum_{j=1}^{N_t}
\log \frac{\bigl\{(2J)^2+(y-q_{tj})^2\bigr\}\bigl\{(2J+2)^2+(y-q_{tj})^2\bigr\}}{\bigl\{(2J+1)^2+(y-q_{tj})^2\bigr\}^2}.
\end{align}
Thus, we end up with
\begin{align}
F_{02}(\{q\})&= \frac{1}{2} \sum_{s,t=2}^{\nu}\sum_{i_s=1}^{N_s}\sum_{j_t=1}^{N_t} 
\sum_{k=1}^{\infty} \left(
 \sum_{J_{st}=kn_1+\frac{n_s-n_t}{2}}^{kn_1+\frac{n_s+n_t}{2}-1}
 -\sum_{J_{st}=kn_1-\frac{n_s+n_t}{2}}^{kn_1-\frac{n_s-n_t}{2}-1}
\right) \nonumber\\
&\qquad \times
\log \frac{\bigl\{(2J_{st})^2+(q_{si_s}-q_{tj_t})^2\bigr\}\bigl\{(2J_{st}+2)^2+(q_{si_s}-q_{tj_t})^2\bigr\}}{\bigl\{(2J_{st}+1)^2+(q_{si_s}-q_{tj_t})^2\bigr\}^2}
+\mathcal{O}\left(e^{-\frac{\pi x_m}{n_1}}\right).
\label{general F02 result}
\end{align}
By substituting $Z(\{q\})=\exp\{F_X+N_1F_{01}(\{q\})+F_{02}(\{q\})+\cdots\}$ with \eqref{general F01 result} and \eqref{general F02 result}
into the eigenvalue integral \eqref{Z general}, 
and taking the double scaling limit \eqref{DSL} of the eigenvalue integral divided by $Z_X$, 
we obtain 
\begin{align}
Z_{\mathrm{DSL}}
 &=\int \prod_{s=2}^{\nu}\prod_{i=1}^{N_s}dq_{si} 
 \times \prod_{s,t=2}^{\nu}\prod_{i_s=1}^{N_s}\prod_{j_t=1}^{N_t}
 \hspace{-5.5mm} {\phantom{\prod}}^{\prime}
 \biggl[
 \prod_{J=\frac{|n_s-n_t|}{2}}^{\frac{n_s+n_t}{2}-1}
 \mathcal{B}(J,q_{si_s}-q_{tj_t}) \nonumber\\
 &\quad \times 
 \prod_{k=1}^{\infty}\biggl[
 \prod_{J_{st}=kn_1+\frac{|n_s-n_t|}{2}}^{kn_1+\frac{n_s+n_t}{2}-1}
 \mathcal{B}(J_{st},q_{si_s}-q_{tj_t})
 \prod_{J_{st}=kn_1-\frac{n_s+n_t}{2}}^{kn_1-\frac{|n_s-n_t|}{2}-1}
 \mathcal{B}(J_{st},q_{si_s}-q_{tj_t})^{-1} \biggr]
 \biggr]
 \nonumber\\
 &\quad \times
 \exp\left\{{-\frac{4}{\pi g_0}\sum_{s=2}^{\nu}\sum_{i_s=1}^{N_s}
 \sin\frac{\pi n_s}{n_1}\cosh\frac{\pi q_{si_s}}{n_1}}\right\}. 
 \label{Z_ns5limit general}
\end{align}

Next, we consider the large $N$ limit,
\begin{align}
N_s\to \infty \quad \text{with} \quad t_s\equiv g_0N_s=\mathrm{fixed}, \qquad (s=2,\cdots,\nu)
\end{align}
and the loop equation for $R^{(s)}(z) \; (s=2,\cdots,\nu)$ as in the previous subsection.
Assuming that $\langle R^{(s)}(z)\rangle_{\mathrm{DSL}}$ has a single branch cut on the interval $[-y_m^{(s)}, y_m^{(s)}]\subset \mathbb{R}$ 
and expanding $\langle R^{(s)}(z)\rangle_{\mathrm{DSL}}$ and $\langle \rho^{(s)}(y)\rangle_{\mathrm{DSL}}$ as in \eqref{rho2 DSL expansion},
one finds that the leading contribution to the loop equation is given by 
\begin{align}
\sum_{t=2}^{\nu}\sum_{k=-\infty}^{\infty} \biggl[
&\frac{1-\mathsf{D}}{1+\mathsf{D}} \left(\mathsf{D}^{|2kn_1+n_s-n_t|} - \mathsf{D}^{|2kn_1+n_s+n_t|} \right)R^{(s)}_{\mathrm{DSL},0}(y+i\epsilon) \nonumber\\
& +\frac{1-\mathsf{D}^{-1}}{1+\mathsf{D}^{-1}}\left(\mathsf{D}^{-|2kn_1+n_s-n_t|} - \mathsf{D}^{-|2kn_1+n_s+n_t|} \right)R^{(s)}_{\mathrm{DSL},0}(y-i\epsilon)
\biggr] 
 -\frac{4}{n_1 t_s}\sin\frac{\pi n_s}{n_1}\sinh\frac{\pi y}{n_1}=0,
\label{ns5 limit loop eq general vacua}
\end{align}
where $y\in [-y_m^{(s)},\: y_m^{(s)}]$.
Moreover, we consider the limit
\begin{align}
&n_1\to\infty, \;\; n_s\to \infty, \;\;  y_m^{(s)}\to\infty \;\: (t_s\to \infty)
 \;\; \text{with} \;\; \frac{n_u}{y_m^{(s)}} \;\; \text{held fixed}. \nonumber\\
 &\qquad (s=2,\cdots,\nu; \; u=1,\cdots, \nu)
\end{align}
This limit corresponds, in the gravity dual, to the regime where the NS5-brane charges are large and a typical scale of the geometry is larger than the string scale,
so that the classical gravity description becomes reliable.
Then, using \eqref{(1-D)/(1+D) expansion}, Eq.~\eqref{ns5 limit loop eq general vacua} reduces to
\begin{align}
\sum_{t=2}^{\nu}\sum_{k=-\infty}^{\infty} \biggl[
& \left(\mathsf{D}^{|2kn_1+n_s-n_t|} - \mathsf{D}^{|2kn_1+n_s+n_t|} \right)R^{(s)}_{\mathrm{DSL},0}(y+i\epsilon) \nonumber\\
& - \left(\mathsf{D}^{-|2kn_1+n_s-n_t|} - \mathsf{D}^{-|2kn_1+n_s+n_t|} \right)R^{(s)}_{\mathrm{DSL},0}(y-i\epsilon)
\biggr] 
 = -2i\mu_s+i\frac{8}{\pi t_s}\sin\frac{\pi n_s}{n_1}\cosh\frac{\pi y}{n_1},
\label{ns5 limit loop eq general vacua 2}
\end{align}
where $\mu_s$ is an integration constant.
In terms of the eigenvalue distributions $\rho^{(s)}_{\mathrm{DSL},0}(y)$, this can be rewritten as
\small
\begin{align}
&\rho^{(s)}_{\mathrm{DSL},0}(y)
- \sum_{t=2}^{\nu}\sum_{k=-\infty}^{\infty} \frac{1}{\pi} \int_{-y^{(t)}_{m}}^{y^{(t)}_{m}} dy' 
\left(
\frac{|2kn_1+n_s+n_t|}{(y-y')^2+|2kn_1+n_s+n_t|^2} - \frac{|2kn_1+n_s-n_t|}{(y-y')^2+|2kn_1+n_s-n_t|^2}\right)\rho^{(t)}_{\mathrm{DSL},0}(y') \nonumber\\
& =\frac{\mu_s}{\pi} - \frac{4}{\pi^2 t_s} \sin\frac{\pi n_s}{n_1} \cosh\frac{\pi y}{n_1},
\label{DSL rhos eq}
\end{align}
\normalsize
where the argument of $\rho^{(s)}_{\mathrm{DSL},0}(y)$ spans the interval $[-y_m^{(s)},y_m^{(s)}]$ 
and $\rho^{(s)}_{\mathrm{DSL},0}(y_m^{(s)})=0$.
The integral equations \eqref{DSL rhos eq} exactly match those determining the gravity dual of a general nontrivial vacuum of LST \cite{Asano:2022brd}.

\section{Summary and discussion} 
\label{summary}

In this paper, we investigated the partition function of PWMM and showed that there exists a nontrivial double scaling limit 
in which the dual geometry of PWMM in the gauge/gravity correspondence reduces to the IIA NS5-brane geometry.
For this purpose, we studied the eigenvalue integral obtained by applying the localization computation \cite{Asano:2012zt} to the partition function 
around a fixed vacuum. The vacuum of PWMM is characterized by a reducible representation of $SU(2)$.
We focused on the largest irreducible representation of dimension $n_1$ with multiplicity $N_1$, and on the corresponding eigenvalues $x_i$.
We then considered the large $N_1$ limit, where $N_1\to\infty$ with $\lambda=g^2N_1$ fixed, and integrating out the eigenvalues $x_i$ 
to obtain an effective theory for remaining eigenvalues.
It was shown that this can be achieved by utilizing the large $N_1$ expansion of the resolvent obtained from the loop equation.
We found that the effective eigenvalue integral indeed admits the double scaling limit,
where the fixed quantity can be identified with the maximum value of the dilaton (effective string coupling) through the comparison with the gravity dual.
Since the localization computation in \cite{Asano:2012zt} is applicable to the expectation value of 
any function of the 1/4 BPS field \eqref{quarter_BPS_field}, 
our result shows the existence of the limit not only for the partition function but also for all correlation functions in the 1/4 BPS sector.
Thus, the resultant eigenvalue integral in the limit is expected to describe the 1/4 BPS sector of IIA LST on $R\times S^5$.

The double scaling limit was first found in \cite{Ling:2006up} on the gravity side and 
then its existence on the gauge theory side was argued based on numerical computation in \cite{Asano:2022brd}.
In contrast, the present study is entirely analytical on the gauge theory side
 and consistency with the gravity side gives evidence for the gauge/gravity correspondence to be true.
Although the double scaling limit on the gravity side suggests that 
the eigenvalue integral \eqref{Z_ns5limit} or \eqref{Z_ns5limit general} describes the 1/4 BPS sector of LST characterized by \eqref{quarter_BPS_field},
it is not clear how the 1/4 BPS field \eqref{quarter_BPS_field} should be interpreted in LST.
Clarifying its meaning will provide a deeper understanding of LST.

It is important to note that we derived the perturbative solution for the resolvent at the leading order in the large $N_1$ expansion.
Specifically, we systematically obtained the solution perturbatively in large $\lambda$ by employing the method developed 
in \cite{Volin:2009tqx,Marino:2019fuy,Reichert:2020ymc}.
Since finite $\lambda$ corrections on the gauge theory side correspond to finite string scale corrections on the gravity side,
it would be interesting to compare our results with the string scale corrections to the gravity dual of PWMM.
Furthermore, we performed the calculation with $n_1$ kept finite, where $n_1$ is the dimension of the largest irreducible representation of $SU(2)$. 
Since $n_1$ corresponds to the NS5-brane charge on the gravity side, which is usually assumed to be large, 
our results may provide new insights into finite NS5-brane charge corrections in the gravity dual.

It is also intriguing to compare our result in subsection \ref{sec:DSL_nu=2} with the work in \cite{Lin:2006tr},
which investigated instanton solutions in the dual gravity background of PWMM as well as LST on $R\times S^5$.
In a perturbative regime where the geometry is regarded as unchanged, the instanton is equivalent to the creation of probe D0-branes 
and its action at weak string coupling is evaluated by considering a Euclidean D2-brane wrapping a nontrivial $S^3$ 
in the geometry \cite{Lin:2005nh,Lin:2006tr}.
The number of the created D0-branes is equal to the $H_3$-flux through the $S^3$ \cite{Maldacena:2001ss,Maldacena:2001xj}.
The instanton action in the NS5-brane background is then calculated as $\frac{2}{g_0}$ \cite{Lin:2006tr}, 
where $g_0$ is the same quantity as in \eqref{DSL}.
Let us compare this with our result \eqref{Z_ns5limit}, which describes $n_2N_2$ D0-branes in the NS5-brane background with $n_1$ $H_3$-flux.
In the weak coupling limit, the action in \eqref{Z_ns5limit} reads $\frac{4N_2}{\pi g_0}\sin\frac{\pi n_2}{n_1}$.
The instanton picture and the probe approximation above are valid when $n_2N_2=n_1\gg n_2$. 
So, we obtain the action $\frac{4N_2}{\pi g_0}\sin\frac{\pi n_2}{n_1}\simeq \frac{4}{g_0}$.
This value is twice the instanton action in the dual gravity.
The difference seems to arise from the following reason.
The eigenvalue integral \eqref{starting eigenvalue integral}, from which we have started, was derived from the partition function for a fixed vacuum at $\tau\to \pm \infty$ 
as explained in section \ref{Some known aspects of the plane wave matrix model}.
Thus, \eqref{Z_ns5limit} should also be regarded as that for a fixed vacuum.
In the current case, there are $n_2N_2$ probe D0-branes in the NS5-brane background at $\tau\to \pm \infty$.
Hence, the action we have evaluated above, $\frac{4}{g_0}$, may be interpreted as 
that of the bounce solution which creates and annihilates the probe branes. 
If this is the case, our action should be twice the instanton action in the weak string coupling limit. 
It is interesting to see if the eigenvalue integral \eqref{Z_ns5limit} and the system of the probe branes in the NS5-background are related in a more rigorous manner.

There are also several other interesting directions worth studying.
Firstly,
it is important to investigate 
the matrix integral we obtained in the double scaling limit 
furthermore
to gain insights into LST.
Secondly,
the existence of a nontrivial double scaling limit was also predicted for other gauge theories 
such as SYM on $R \times S^2$ and SYM on $R \times S^3/Z_k$ \cite{Ling:2006xi}.
Similar analysis in terms of the eigenvalue integral should be possible for these theories \cite{Asano:2014eca}. 
Lastly,
it is intriguing to explore whether an analogous double scaling limit exists in the eigenvalue integral for the polarized IKKT matrix model \cite{Bonelli:2002mb},
which is a supersymmetric mass-deformation of the IKKT matrix model \cite{Ishibashi:1996xs}
and shares various distinctive features with PWMM \cite{Hartnoll:2024csr,Komatsu:2024bop,Komatsu:2024ydh,Hartnoll:2025ecj,Chou:2025rwy}.
We hope to reveal a formulation of LST
through such analysis.

\section*{Acknowledgments}
This work was supported by JSPS KAKENHI Grant Numbers JP23K03405 and JP24K07036.

\appendix

\section{Solutions of \eqref{Loop eq leading} and  \eqref{Loop eq leading approx} in the strong coupling expansion}
\label{app:detail_solution}

We examine the eigenvalue integral \eqref{Z nu=1} for $\nu=1$ in detail.
We are interested in the regime defined by taking the large $N$ limit, where $N\to \infty$ with $\lambda=g^2N$ held fixed, and subsequently taking the large $\lambda$ limit.
The eigenvalue integral in the large $N$ limit is governed by the resolvent $R_0(x)$, 
which is subject to \eqref{Loop eq leading}. 
The large $\lambda$ limit corresponds to the regime where the eigenvalue extent $x_m$ becomes large.
If we also consider the large $n$ limit with $\frac{n}{x_m}$ held fixed, the resolvent satisfies \eqref{Loop eq leading approx}.
Although solving \eqref{Loop eq leading} or \eqref{Loop eq leading approx} may seem challenging, 
a perturbative expansion in the large $\lambda$ limit has been developed for similar equations \cite{Volin:2009tqx,Marino:2019fuy,Reichert:2020ymc}.
We will apply their method to our case. 

\subsection{Preliminary}

In this appendix, for simplicity, we use the following notation,
\begin{align}
R(z)&= \frac{\pi\lambda}{2nx_m^2}R_0(x_mz), \label{R and R0} \\
\kappa&= \frac{2n}{x_m}, \label{kappa_xm} \\
\tilde{\mu}&= \frac{\lambda}{2nx_m^2}\mu. \label{mutilde_mu}
\end{align}
The resolvent $R(z)$ possesses a single branch cut on the interval $[-1,1]$.
The eigenvalue distribution for the resolvent is denoted by $f(x)$,
\begin{align}
R(z)=\int_{-1}^{1} dy \frac{f(y)}{z-y}. \label{def R}
\end{align}
We define the large $z$ expansion of the resolvent $R(z)$ by
\begin{align}
R(z)=\sum_{a=0}^{\infty}T_a(\kappa)z^{-a-1},
\end{align}
where the coefficients $T_a(\kappa)$ are the moments of $f(x)$,
\begin{align}
T_a(\kappa)=\int_{-1}^{1}dx\, x^a f(x). \label{Ta}
\end{align}
From $R_0(z)\to \frac{1}{z}$, $R(z)\to \frac{1}{z}T_0(\kappa)$ $(z\to \infty)$ and \eqref{R and R0},
the relation between $\lambda$ and $x_m$ is obtained as
\begin{align}
\lambda=\frac{2nx_m^3}{\pi}T_0(\tfrac{2n}{x_m}) \label{lambda_xm}
\end{align}

\subsection{Solution of \eqref{Loop eq leading approx}} 
\label{solution of fredholm eq}

First, we begin by solving \eqref{Loop eq leading approx}, which is technically simpler, 
and then turn to \eqref{Loop eq leading} in Appendix \ref{app:solution of the full equation}.
In what follows, quantities associated with \eqref{Loop eq leading approx} will be denoted with a bar.

We consider
\begin{align}
(1-\mathsf{D}^{\kappa}) \bar{R}(x+i0) - (1-\mathsf{D}^{- \kappa})\bar{R}(x-i0) &= -2\pi i (\tilde{\mu} - x^2) \quad (x \in [-1,1]),
\label{fredholm eq rescale}
\end{align}
which is obtained from \eqref{Loop eq leading approx} with \eqref{R and R0}, \eqref{kappa_xm} and \eqref{mutilde_mu}.
The eigenvalue distribution $\bar{f}(x)$ obeys
\begin{align}
\bar{f}(x)-\frac{1}{\pi}\int_{-1}^{1}dy\frac{\kappa}{(x-y)^2+\kappa^2}\bar{f}(y)=\tilde{\mu}-x^2. \label{fredholm eq rescale f}
\end{align}
$\tilde{\mu}$ is determined by the condition $\bar{f}(\pm 1)=0$.

Let us write the solution as 
\begin{align}
\bar{R}(z)=-\bar{R}^{[2]}(z)+\left(\tilde{\mu}-\frac{1}{2}-\frac{\kappa^2}{3}\right)\bar{R}^{[0]}(z),
\label{R_from_R0_and_R2 fredholm}
\end{align}
where $\bar{R}^{[r]}(z)$ ($r=0,2$) are defined as the resolvents that have a discontinuity along the interval $[-1,1]$ 
and satisfy for $x\in [-1,1]$
\begin{align}
(1-\mathsf{D}^{\kappa}) \bar{R}^{[0]}(x+i0) - (1-\mathsf{D}^{- \kappa})\bar{R}^{[0]}(x-i0) &= -2\pi i, \label{R0 eq fredholm} \\
(1-\mathsf{D}^{\kappa}) \bar{R}^{[2]}(x+i0) - (1-\mathsf{D}^{- \kappa})\bar{R}^{[2]}(x-i0) 
&= -2\pi i \left(z^2-\frac{1}{2}-\frac{\kappa^2}{3}\right). \label{R2 eq fredholm}
\end{align}
The constant term in the right-hand side of \eqref{R2 eq fredholm} has been added for convenience in the calculation.
We decompose the resolvent into the regular part and the discontinuous part,  $\bar{R}^{[r]}(z)=\bar{R}^{[r]}_{\mathrm{reg}}(z)+\bar{R}^{[r]}_{\mathrm{disc}}(z)$  ($r=0,2$).
From \eqref{R0 eq fredholm} and \eqref{R2 eq fredholm}, the regular part is obtained as
\begin{align}
\bar{R}^{[0]}_{\mathrm{reg}}(z)&=\frac{\pi}{\kappa}z, \label{R[0]_reg fredholm} \\
\bar{R}^{[2]}_{\mathrm{reg}}(z)&=\frac{\pi}{\kappa}\Big(\frac{1}{3}z^3-\frac{1}{2}z\Big), \label{R[2]_reg fredholm}
\end{align}
The discontinuous part then obeys
\begin{align}
(1-\mathsf{D}^{\kappa})\bar{R}^{[r]}_{\mathrm{disc}}(x+i0)-(1-\mathsf{D}^{- \kappa})\bar{R}^{[r]}_{\mathrm{disc}}(x-i0)&=0,
\label{fredholm eq rescale disc r} 
\end{align}
and 
\begin{align}
\bar{R}^{[r]}_{\mathrm{disc}}(z)\to -\bar{R}^{[r]}_{\mathrm{reg}}(z) + \mathcal{O}\big(\tfrac{1}{z}\big) \quad \mathrm{in} \; z\to \infty.
\label{R_asymptotic fredholm}
\end{align}
We denote the eigenvalue distributions corresponding to $\bar{R}^{[r]}(z)$ by $\bar{f}^{[r]}(z)$.
The condition $\bar{f}(\pm 1)=0$ is fulfilled if $\tilde{\mu}$ satisfies
\begin{align}
\tilde{\mu}=\frac{\bar{f}^{[2]}(1)}{\bar{f}^{[0]}(1)}+\frac{1}{2}+\frac{\kappa^2}{3}.
\label{tildemu_condition}
\end{align}

We are in particular interested in the parameter region where $x_m$ is large. This implies $\kappa\ll 1$.
To analyze \eqref{fredholm eq rescale disc r} in this regime, we employ the systematic method developed in \cite{Volin:2009tqx,Marino:2019fuy,Reichert:2020ymc},
where similar types of equations are perturbatively solved\footnote{
In fact,  Eq.~\eqref{R0 eq fredholm}, when rewritten in terms of $\bar{f}^{[0]}(x)$, is equivalent to the Lieb-Liniger integral equation \cite{Lieb:1963rt} and is analyzed in \cite{Marino:2019fuy,Reichert:2020ymc}.
}.
Specifically, perturbative expansion in $\kappa$ is considered in two regimes,
\begin{align}
\text{bulk regime:}&\quad \kappa\to 0, \quad \text{and} \quad x=\text{fixed}, \label{bulk regime fredholm} \\
\text{edge regime:}&\quad \kappa\to 0, \quad x\to 1 \quad \text{with} \quad t\equiv 2\frac{x-1}{\kappa}=\text{fixed}, \label{edge regime fredholm}
\end{align}
and all the expansion coefficients of each regime are determined from a consistency of the two expansions.
Below, we consider solving \eqref{fredholm eq rescale disc r} for $\bar{R}^{[r]}_{\mathrm{disc}}(z)$ ($r=0,2$) in these regimes.


\subsubsection{Solving the discontinuous part of $\bar{R}^{[r]}(z)$}
\label{Solving the discontinuous part of R[r](z)}

We first consider the bulk regime \eqref{bulk regime fredholm}. 
In this regime, the derivative expansion $\mathsf{D}=1+i\partial +\cdots$ can be applied.
We thus look for the solution to \eqref{fredholm eq rescale disc r} and \eqref{R_asymptotic fredholm} with the following ansatz for $r=0$ and $r=2$,
\begin{align}
\bar{R}_{\text{disc,b}}^{[0]}(x)
&=-\frac{\pi}{\kappa}(x^2-1)^{\frac{1}{2}}
+\sum_{l,m=0}^{\infty}\sum_{k=0}^{l+m+1}\bar{c}^{[0]}_{lmk}\kappa^{l+m}\frac{x^{p_0(k)}}{(x^2-1)^{l+\frac{1}{2}}}
\log^k\Big(\frac{x-1}{x+1}\Big),  \nonumber\\
\bar{R}_{\text{disc,b}}^{[2]}(x)
&=-\frac{\pi}{3 \kappa}(x^2-1)^{\frac{3}{2}}
+\sum_{l,m=0}^{\infty}\sum_{k=0}^{l+m+1}\bar{c}^{[2]}_{lmk}\kappa^{l+m}\frac{x^{p_0(k)}}{(x^2-1)^{l-\frac{1}{2}}}
\log^k\Big(\frac{x-1}{x+1}\Big).
\label{expansion_bulk_0_2 fredholm}
\end{align}
Here, $p_0(k)=0\; (\text{$k$ even}),\; 1\; (\text{$k$ odd})$.
The coefficients $\bar{c}^{[r]}_{lmk}$ are polynomials of $\log(\kappa)$ and determined by comparing with the edge regime expansion
we describe below.
Note that when $r=2$, for \eqref{R_asymptotic fredholm} to hold, the coefficient $\bar{c}^{[2]}_{lmk}$ must satisfy the condition;
\begin{align}
\bar{c}^{[2]}_{0m0}=2\bar{c}^{[2]}_{0m1}.
\end{align}

In the edge regime, we focus on 
\begin{align}
\bar{R}_{\text{disc,e}}^{[r]}(t)\equiv \bar{R}_{\text{disc}}^{[r]}\left(1+\frac{\kappa t}{2}\right) \quad \text{for} \quad  \kappa\ll 1,
\label{R0_edge_disc}
\end{align}
which has a branch cut along the negative real axis and satisfies, for $t\leq 0$,
\begin{align}
(1-\mathsf{D}^{2}) \bar{R}^{[r]}_{\mathrm{disc,e}}(t+i0)-(1-\mathsf{D}^{-2}) \bar{R}^{[r]}_{\mathrm{disc,e}}(t-i0)=0.
\label{fredholm eq rescale disc 0 edge} 
\end{align}
From the small-$|t|$ expansion of \eqref{fredholm eq rescale disc 0 edge},
one finds that $\bar{R}^{[r]}_{\mathrm{disc,e}}(t+i0)-\bar{R}^{[r]}_{\mathrm{disc,e}}(t-i0)$ admits a power series expansion in $t^a$ with non-negative integer $a$.
Consequently, for $|t|\ll 1$, $\bar{R}^{[r]}_{\mathrm{disc,e}}(t)$ can be written as
\begin{align}
\bar{R}_{\text{disc,e}}^{[r]}(t)\big|_{|t|\ll 1} \simeq -\log(t) \times \sum_{a=0}^{\infty} \bar{f}_a^{[r]} t^a, \label{small t expansion of edgeR fredholm}
\end{align}
where $\bar{f}_a^{[r]}$ are the expansion coefficients.
We rewrite $\bar{R}_{\text{disc,e}}^{[r]}(t)$ in terms of the Laplace transform, $\hat{\bar{R}}_{\text{disc,e}}^{[r]}(s)$,
\begin{align}
\bar{R}_{\text{disc,e}}^{[r]}(t)=\int_0^\infty ds\, e^{-st} \hat{\bar{R}}_{\text{disc,e}}^{[r]}(s),
\label{Laplace_transform}
\end{align}
then \eqref{fredholm eq rescale disc 0 edge} becomes
\begin{align}
(1-e^{-2is})\hat{\bar{R}}_{\text{disc,e}}^{[r]}(s-i0)-(1-e^{2is})\hat{\bar{R}}_{\text{disc,e}}^{[r]}(s+i0)=0,
\label{fredholm eq_R0_edge_s_disc}
\end{align}
for $s<0$.
The edge-regime expansion of $\hat{\bar{R}}_{\text{disc,e}}^{[r]}(s)$ can be determined from its analytic property, which is summarized as follows:
order by order in the $\kappa$ expansion, it is analytic except on the negative real axis and 
 at large $s$  it has a $\frac{1}{s^a}$-expansion with positive integer $a$, which follows from \eqref{small t expansion of edgeR fredholm} 
 with the (analytically continued) Laplace transform
\begin{align}
s^{-a}\leftrightarrow \frac{(-1)^a}{(a-1)!}\log z \; z^{a-1}.
\end{align}
These properties as well as the structure of poles and zeros  should be consistent with \eqref{fredholm eq_R0_edge_s_disc}.
Thus, the edge-regime ansatz for $\hat{\bar{R}}_{\text{disc,e}}^{[r]}(s)$ can be taken as\footnote{
The discontinuity for $s<0$ shown in \eqref{fredholm eq_R0_edge_s_disc} indicates the presence of $s^{-1/2}\exp\left[\frac{s}{\pi}\log\left(\frac{\pi e}{s}\right)\right]$.
For the $1/s$-expansion to be valid at large $s$, it must come with $\Gamma(\tfrac{s}{\pi}+1)$.
In \eqref{fredholm eq_R0_edge_s_disc}, the poles of $\Gamma(\tfrac{s}{\pi}+1)$ are canceled by the zeros of $(1-e^{\pm 2is})$.
}
\begin{align}
\hat{\bar{R}}_{\text{disc,e}}^{[r]}(s)=
\frac{1}{\sqrt{\kappa}}
\bar{\Phi}(s)
\bar{Q}^{[r]}(s)
\label{expansion_edge_0 fredholm}
\end{align}
where 
\begin{align}
\bar{\Phi}(s)&\equiv \frac{1}{s^{\frac{3}{2}}}\exp\left[\frac{s}{\pi}\log\left(\frac{\pi e}{s}\right)\right]\Gamma\left(\frac{s}{\pi}+1\right), \label{Phi(s) fredholm} \\
\bar{Q}^{[r]}(s)&\equiv \sum_{l,m=0}^{\infty}\bar{Q}^{[r]}_{lm}\frac{\kappa^{l+m}}{s^l}. \label{Qr(s) fredholm}
\end{align}
The coefficients $\bar{Q}^{[r]}_{lm}$ are polynomials of $\log(\kappa)$.
Note that the gamma function $\Gamma(\frac{s}{\pi}+1)$ has simple poles at $s=-\pi,-2\pi,\cdots$ and admits the $1/s$-expansion in $s\to \infty$ as
\begin{align}
\Gamma\left(\frac{s}{\pi}+1\right)&\sim \sqrt{2s}\exp\left[\frac{s}{\pi}\log\frac{s}{\pi e}\right] \left(1 + \mathcal{O}\left(\frac{1}{s}\right)\right).
\end{align}

The coefficients $\bar{Q}^{[r]}_{lm}$ and $\bar{c}^{[r]}_{lmk}$ can be fixed simultaneously by matching the bulk expansion \eqref{expansion_bulk_0_2 fredholm} 
and the edge expansion \eqref{expansion_edge_0 fredholm} as follows:
In the bulk expansion \eqref{expansion_bulk_0_2 fredholm}, we put $x=1+\kappa t/2$ and take the $\kappa\to 0$ limit,
while in the edge expansion \eqref{expansion_edge_0 fredholm}, we expand it around $s=0$ and perform the Laplace transform
\eqref{Laplace_transform}. 
By comparing the coefficients of  $\kappa, t, \log(t)$, 
both $\bar{Q}^{[r]}_{lm}$ and $\bar{c}^{[r]}_{lmk}$ can be determined as functions of $\log(\kappa)$.

The coefficients for $r=0$, $\bar{Q}^{[0]}_{lm}$ and $\bar{c}^{[0]}_{lmk}$,  are obtained as
\begin{align}
& \bar{Q}^{[0]}_{00} = \frac{\sqrt{\pi}}{2}, \quad 
 \bar{Q}^{[0]}_{01} = - \frac{L}{8 \sqrt{\pi}}, \quad
 \bar{Q}^{[0]}_{02} = - \frac{L^{2}}{64 \pi^{\frac{3}{2}}} - \frac{L}{16 \pi^{\frac{3}{2}}}, \quad
 \bar{Q}^{[0]}_{10} = - \frac{3\sqrt{\pi}}{32}, \nonumber\\
& \bar{Q}^{[0]}_{11} =  - \frac{3 L}{128 \sqrt{\pi}} - \frac{5}{128 \sqrt{\pi}}, \quad
 \bar{Q}^{[0]}_{20} = - \frac{15\sqrt{\pi}}{1024}, \quad
\nonumber \\[3mm]
& \bar{c}^{[0]}_{000} = - \frac{L}{2} + \frac{1}{2}, \quad
 \bar{c}^{[0]}_{001} = \frac{1}{2}, \quad
 \bar{c}^{[0]}_{010} = \frac{L^{2}}{8 \pi} - \frac{1}{4 \pi}, \quad
 \bar{c}^{[0]}_{011} = 0, \nonumber\\
& \bar{c}^{[0]}_{012} = 0, \quad
 \bar{c}^{[0]}_{100} = \frac{L^{2}}{8 \pi} + \frac{L}{4 \pi} - \frac{3}{8 \pi} + \frac{\pi}{12}, \quad
 \bar{c}^{[0]}_{101} = - \frac{L}{4 \pi} - \frac{1}{4 \pi}, \quad
 \bar{c}^{[0]}_{102} = \frac{1}{8 \pi}, 
\label{coeff r=0 fredholm}
\end{align}
where $L\equiv \log\left(\frac{\kappa}{16\pi}\right)$.
The coefficients for $r=2$, $\bar{Q}^{[2]}_{lm}$ and $\bar{c}^{[2]}_{lmk}$, are 
\begin{align}
& \bar{Q}^{[2]}_{00} = 0,\quad
 \bar{Q}^{[2]}_{01} = - \frac{L}{4 \sqrt{\pi}} - \frac{3}{4 \sqrt{\pi}}, \quad
 \bar{Q}^{[2]}_{02} = \frac{L^{2}}{8 \pi^{\frac{3}{2}}} + \frac{7 L}{16 \pi^{\frac{3}{2}}} - \frac{\sqrt{\pi}}{12} - \frac{3}{8 \pi^{\frac{3}{2}}}, \quad
 \bar{Q}^{[2]}_{10} = -\frac{\sqrt{\pi}}{4}, \nonumber\\
& \bar{Q}^{[2]}_{11} = \frac{15 L}{64 \sqrt{\pi}} + \frac{25}{64 \sqrt{\pi}}, \quad
 \bar{Q}^{[2]}_{20} = \frac{15\sqrt{\pi}}{64}, \quad
\nonumber\\[3mm]
& \bar{c}^{[2]}_{000} = 1, \quad
 \bar{c}^{[2]}_{001} = \frac{1}{2}, \quad
 \bar{c}^{[2]}_{010} = 0, \quad
 \bar{c}^{[2]}_{011} = 0, \nonumber\\
& \bar{c}^{[2]}_{012} = - \frac{1}{4\pi}, \quad
 \bar{c}^{[2]}_{100} = \frac{L^{2}}{8 \pi} + \frac{3 L}{4 \pi} - \frac{7}{8 \pi} - \frac{\pi}{12}, \quad
 \bar{c}^{[2]}_{101} = -\frac{1}{\pi}, \quad
 \bar{c}^{[2]}_{102} = -\frac{1}{8\pi}.
\label{coeff r=2 fredholm}
\end{align}

\subsubsection{Edge behavior of $f^{[r]}(x)$}

Eq.~\eqref{small t expansion of edgeR fredholm} indicates that 
the eigenvalue distribution in the edge regime, $\bar{f}^{[r]}_{\text{e}}(t)=\bar{f}^{[r]}(1+\frac{\kappa t}{2})$ with $\kappa\to 0$, takes the form for $|t|\ll 1$ as
\begin{align}
\bar{f}^{[r]}_{\text{e}}(t)\big|_{|t|\ll 1}\simeq \sum_{a=0}^{\infty} \bar{f}_a^{[r]} t^a. \label{expansion_fr_edge fredholm}
\end{align}

We will express the coefficient $\bar{f}_a^{[r]}$ in terms of $\bar{Q}_{lm}^{[r]}$.
For this, let us consider the $\frac{1}{s}$-expansion of the edge-regime ansatz \eqref{expansion_edge_0 fredholm} around $s=\infty$. 
We denote by $\bar{g}_k$ the coefficients of the large-$s$ expansion of \eqref{Phi(s) fredholm},
\begin{align}
\bar{\Phi}(s)\bigr|_{s\gg 1}=\sum_{k=0}^\infty \frac{\bar{g}_k}{s^{k+1}}
 \label{expansion_gamma fredholm}
\end{align}
The coefficients $\bar{g}_k\; (k=0,1,2,\cdots)$ are given by
\begin{align}
\bar{g}_0&=\sqrt{2},\;\; 
\bar{g}_1=\frac{\sqrt{2}\pi}{12}, \;\;
\bar{g}_2=\frac{\sqrt{2}\pi^2}{288}, \;\;
\bar{g}_3=- \frac{139\sqrt{2}\pi^{3}}{51840}, \;\; \cdots.
\end{align}
We also define
\begin{align}
\bar{Q}^{[r]}_{(a)}
&\equiv \sum_{l=0}^{a}\sum_{m=0}^{\infty} \bar{g}_{a-l}\kappa^{l+m} \bar{Q}^{[r]}_{lm} 
=\bar{g}_a\bar{Q}^{[r]}_{00}+\left(\bar{g}_a\bar{Q}^{[r]}_{01} + \bar{g}_{a-1} \bar{Q}^{[r]}_{10}\right)\kappa +\cdots. 
\label{tildeQn fredholm}
\end{align}
Then, the edge-regime ansatz \eqref{expansion_edge_0 fredholm} around $s=\infty$ can be written as
\begin{align}
\hat{\bar{R}}_{\text{disc,e}}^{[r]}(s)\Bigr|_{s\gg 1}
\simeq \frac{1}{\sqrt{\kappa}}\sum_{a=0}^\infty \frac{\bar{Q}^{[r]}_{(a)}}{s^{a+1}},
\label{Rs_tildeQ fredholm}
\end{align}
By performing the (analytically continued) Laplace transform
and comparing with \eqref{small t expansion of edgeR fredholm}, one finds
\begin{align}
\bar{f}_a^{[r]}=\frac{1}{\sqrt{\kappa}}\frac{(-1)^{a}\bar{Q}^{[r]}_{(a)}}{a!}.
\label{fbar_a}
\end{align}

\subsubsection{Result for $\bar{f}(x)$ in the strong coupling expansion}

Now, we are able to obtain the perturbative expression of $\bar{f}(x)$ as well as $\bar{R}(z)$ at arbitrary order in $\kappa$. 
Below, we present them up to the next-to-leading order.

First, we evaluate $\tilde{\mu}$ in \eqref{tildemu_condition}.
It follows from \eqref{expansion_fr_edge fredholm}, \eqref{fbar_a} and \eqref{tildeQn fredholm} 
that 
$\bar{f}^{[r]}(1)=\frac{\sqrt{2}}{\sqrt{\kappa}}(\bar{Q}^{[r]}_{00}+\bar{Q}^{[r]}_{01}\kappa+\bar{Q}^{[r]}_{02}\kappa^2+\cdots)$
for $r=0,2$.
By using \eqref{coeff r=0 fredholm} and \eqref{coeff r=2 fredholm}, 
$\tilde{\mu}$ can be calculated as
\begin{align}
\tilde{\mu}
&=\frac{1}{2}+\kappa \left(- \frac{\log\left(\frac{\kappa}{16\pi}\right)}{2 \pi} - \frac{3}{2 \pi} \right)
+ \kappa^{2} \left(
\frac{\log\left(\frac{\kappa}{16\pi}\right)^{2}}{8 \pi^{2}} + \frac{\log\left(\frac{\kappa}{16\pi}\right)}{2 \pi^{2}} + \frac{1}{6} - \frac{3}{4 \pi^{2}}
\right) 
+ \mathcal{O}\left(\kappa^{3}(\log\kappa)^3\right).
\label{gamma_solution}
\end{align}

\subsubsection*{$\bullet$ resolvent $\bar{R}(z)$}

The regular part of $\bar{R}(z)$ is obtained from \eqref{R_from_R0_and_R2 fredholm}, \eqref{R[0]_reg fredholm} and \eqref{R[2]_reg fredholm} as
\begin{align}
\bar{R}_{\mathrm{reg}}(z)
= \frac{\pi}{\kappa}\left(-\frac{1}{3}z^3+\left(\tilde{\mu}-\frac{\kappa^2}{3}\right)z.
\right)
\end{align}

The discontinuous part in the bulk regime is given by
\begin{align}
\bar{R}_{\text{disc,b}}(z)
&=
\frac{\pi}{3\kappa} (z^2-1)^{\frac{3}{2}} 
+\left(\frac{\log\left(\frac{\kappa}{16\pi}\right)}{2} + \frac{1}{2} \right)\sqrt{z^2-1}
-\frac{1}{2}z\sqrt{z^2-1}\ln\left(\frac{z-1}{z+1}
\right)
+\mathcal{O}(\kappa(\log\kappa)^2).
\end{align}
The discontinuous part in the edge regime, $\bar{R}_{\text{disc,e}}(t)=\bar{R}_{\text{disc}}(1+\frac{\kappa t}{2})$, is given by
\begin{align}
&\bar{R}_{\text{disc,e}}(t)=\int_0^{\infty} ds\, e^{-ts}\hat{\bar{R}}_{\text{disc,e}}(s),
\end{align}
with 
\begin{align}
\hat{\bar{R}}_{\text{disc,e}}(s)
&= 
\frac{1}{\sqrt{\kappa}}\left[
 \frac{\sqrt{\pi}}{4}\frac{\kappa}{s} 
+ \Bigl(- \frac{3 \log\left(\frac{\kappa}{16\pi}\right)}{16 \sqrt{\pi}} - \frac{1}{4 \sqrt{\pi}}\Bigr)\frac{\kappa^2 }{ s } 
- \frac{15 \sqrt{\pi}}{64}\frac{\kappa }{ s^2 }
\right]
\bar{\Phi}(s) +\mathcal{O}(\kappa^3(\log\kappa)^2) .
\end{align}
For $|t|\ll 1$, it reduces to
\begin{align}
\bar{R}_{\text{disc,e}}(t)|_{|t|\ll 1}&=-\log t\times \left(\bar{f}_1t+\bar{f}_2t^2+\cdots\right),
\end{align}
where 
\begin{align}
\bar{f}_1&=
-\frac{ \sqrt{2\pi} }{4}\sqrt{\kappa}
+ \sqrt{2} \left( 
 \frac{3 \log\left(\frac{\kappa}{16\pi}\right)}{16 \sqrt{\pi}} + \frac{1}{4 \sqrt{\pi}} 
\right)\kappa^{\frac{3}{2}} + \mathcal{O}\left(\kappa^\frac{5}{2}(\log\kappa)^2\right), \nonumber\\
\bar{f}_2&=
 \frac{\sqrt{2} \pi^{\frac{3}{2}}}{96} \sqrt{\kappa}
+ \sqrt{2\pi} \left(
 -\frac{\log\left(\frac{\kappa}{16\pi}\right)}{128} - \frac{49}{384} 
\right)\kappa^{\frac{3}{2}} +\mathcal{O}\left(\kappa^\frac{5}{2}(\log\kappa)^2\right), 
\label{barf1_and_barf2 fredholm}
\end{align}

\subsubsection*{$\bullet$ eigenvalue distribution $\bar{f}(x)$}

The eigenvalue distribution in the bulk regime is given by
\begin{align}
\bar{f}_{\mathrm{b}}(x)
&=\frac{1}{3 \kappa}(1-x^2)^{\frac{3}{2}}
-\frac{1}{\pi}\left(\frac{\log\left(\frac{\kappa}{16\pi}\right)}{2} + \frac{1}{2}  \right)\sqrt{1-x^2}
+\frac{1}{2\pi}x\sqrt{1-x^2}\ln\left(\frac{1-x}{1+x}\right)
+\mathcal{O}\big(\kappa(\log\kappa)^2\big). \label{f_bulk_solution fredholm} 
\end{align}
The eigenvalue distribution in the edge regime, $\bar{f}_{\mathrm{e}}(t)=\bar{f}(1+\frac{\kappa t}{2})$, is given by
\begin{align}
\bar{f}_{\mathrm{e}}(t)|_{|t|\ll 1}&= \bar{f}_1t+\bar{f}_2t^2+\cdots, \label{f_edge_solution fredholm} 
\end{align}
where $\bar{f}_1$ and $\bar{f}_2$ are provided in \eqref{barf1_and_barf2 fredholm}.

\subsubsection*{$\bullet$ the extent of the eigenvalues $x_m$}

Finally, we present $x_m$ as a function of $\lambda$.
From the large $z$ expansion of the resolvent in the bulk regime, we obtain $\bar{T}_0(\kappa)$ in \eqref{Ta} as
\begin{align}
\bar{T}_0(\kappa)=\frac{\pi}{8\kappa}-\frac{\log\left(\frac{\kappa}{16\pi}\right)}{4}-\frac{5}{12}+\mathcal{O}(\kappa(\log\kappa)^2).
\label{T0_solution fredholm}
\end{align}
By using \eqref{lambda_xm} and \eqref{T0_solution fredholm}, 
$\lambda$ can be expressed as a function of $\kappa$.
Then, with \eqref{kappa_xm}, $\bar{x}_m$ can be obtained as a function of $\lambda$,
\begin{align}
\bar{x}_m=(8\lambda)^{\frac{1}{4}}+\frac{n}{\pi}\left(-\log\frac{8\pi(8\lambda)^{\frac{1}{4}}}{n}+\frac{5}{3}\right)+\mathcal{O}\left(\lambda^{-\frac{1}{4}}(\log\lambda)^2\right).
\end{align}

\begin{figure}[htbp]
\begin{center}
\includegraphics[height=60mm]{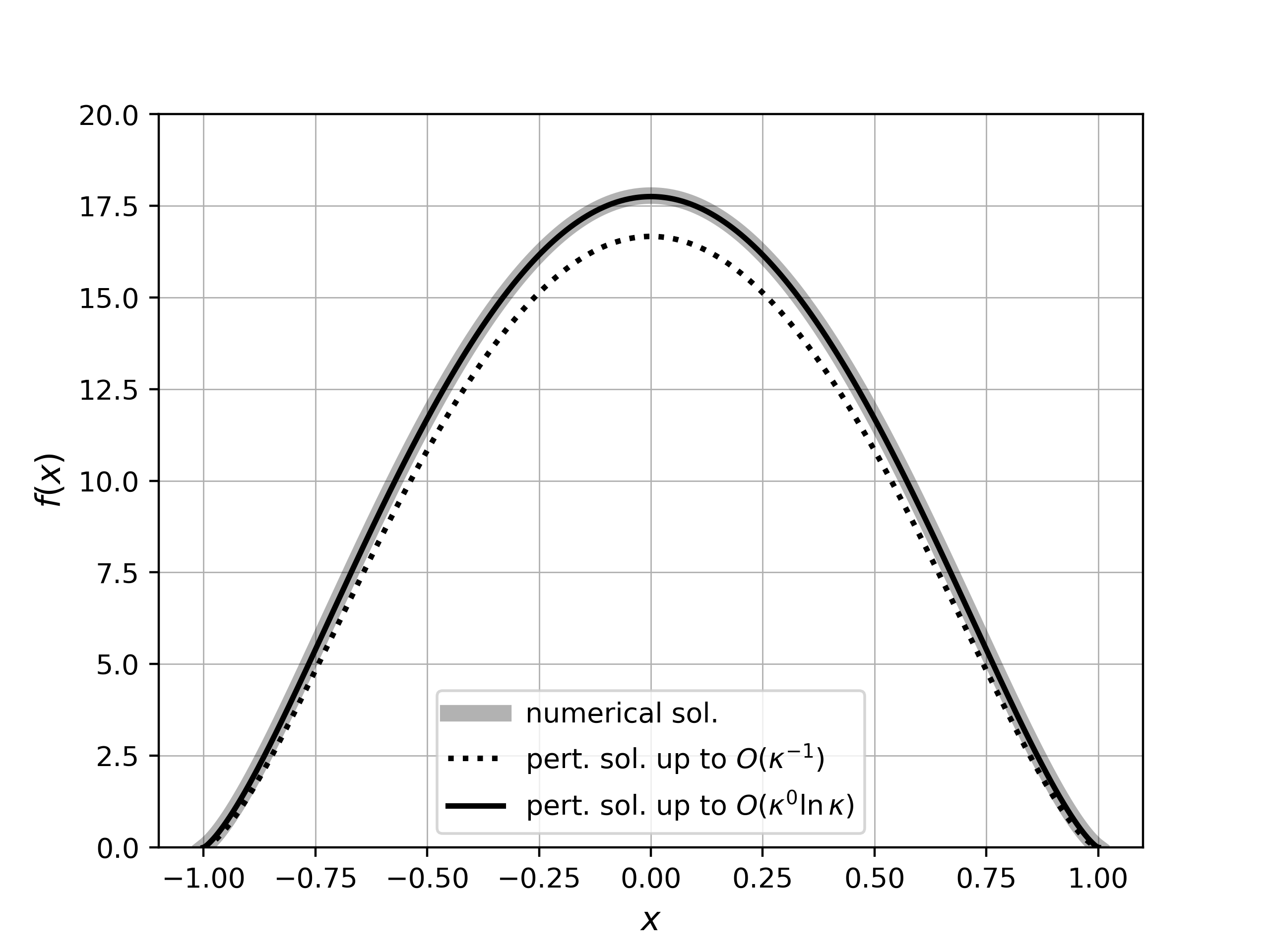}
\includegraphics[height=60mm]{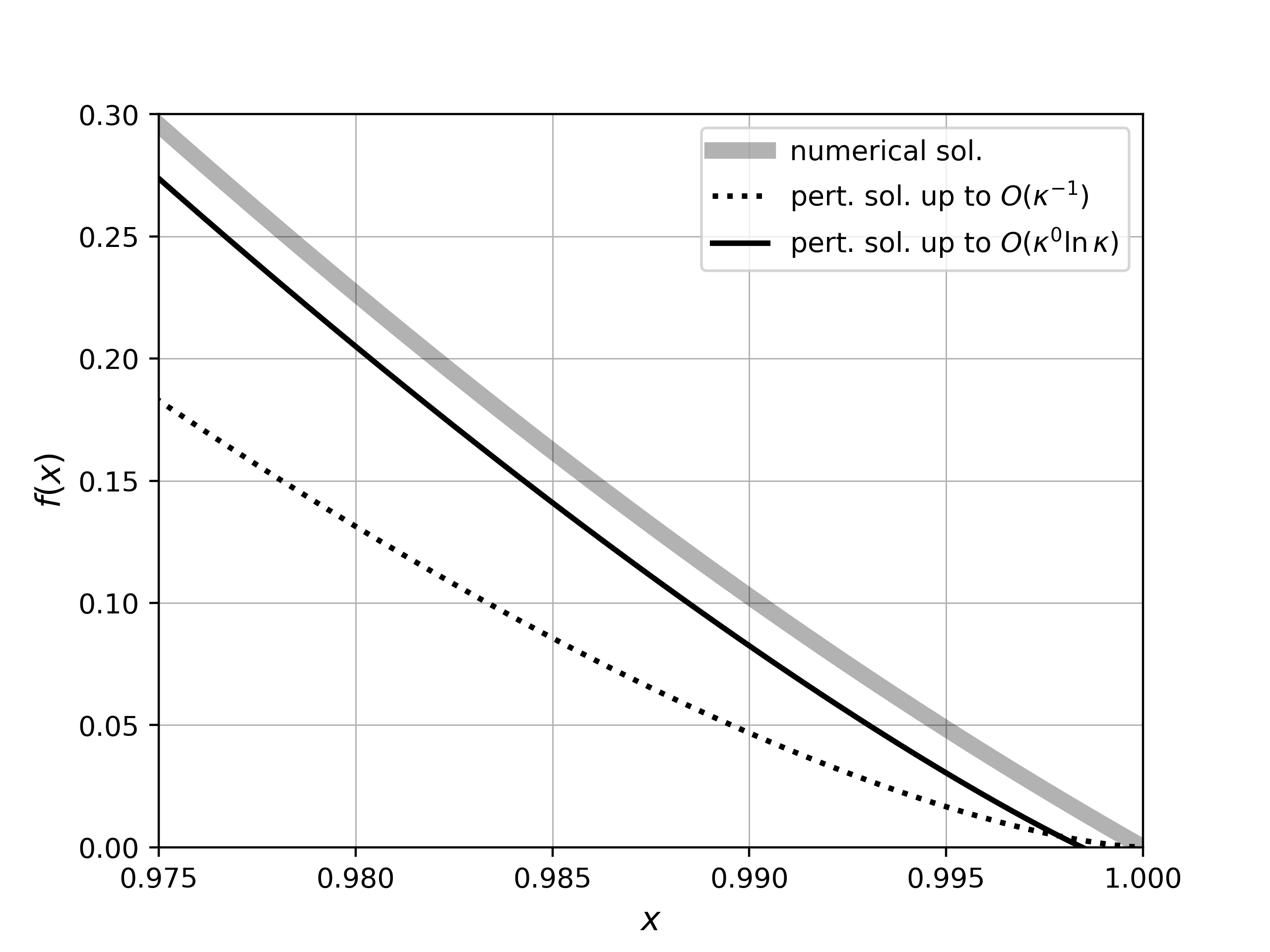}
\caption{
The left figure displays the entire range of the eigenvalue distribution $\bar{f}(x)$ for $\kappa=0.02$,
while the right figure provides the zoom-in of the edge region ($x\sim 1$).
The thick gray solid line represents the numerical solution, which is sufficiently precise to be considered exact.
The black dotted and solid lines correspond to the perturbative bulk-regime solutions \eqref{f_bulk_solution}
up to $\mathcal{O}(\kappa^{-1})$ (leading order) and $\mathcal{O}(\kappa^{0}\log \kappa)$, respectively.
As expected, the latter solution fits the numerical result better than the former away from the edge, 
while both solutions deviate from the numerical result significantly near the edge.
}\label{fig:bulk_solution}
\end{center}

\medskip

\begin{center}
\includegraphics[height=60mm]{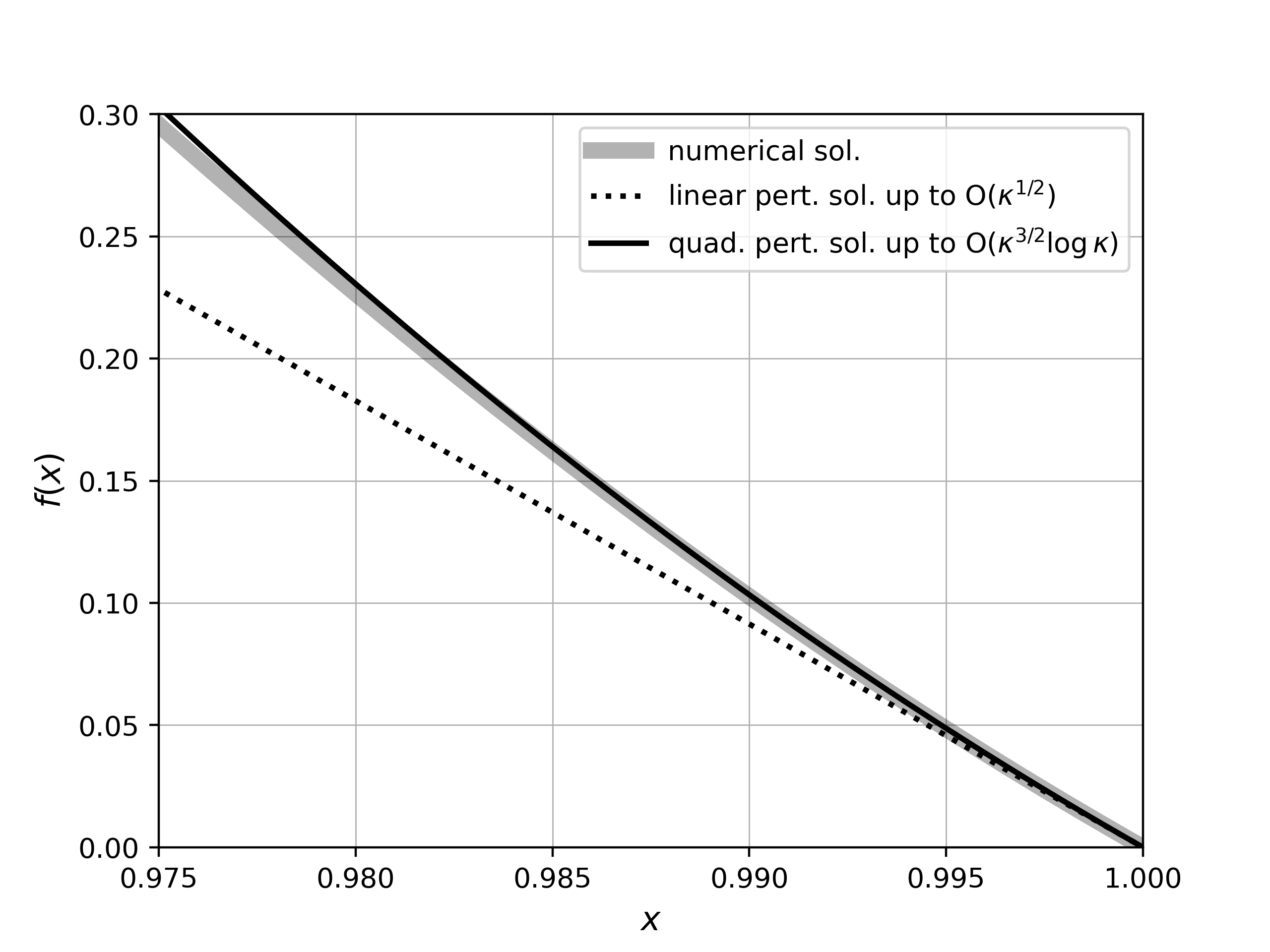}
\caption{
The figure shows the perturbative edge-regime solutions of the eigenvalue distribution $f(x)$ for $\kappa=0.02$.
The thick gray solid line represents the numerical solution, which is precise enough to be considered exact.
The black dotted line corresponds to the solution \eqref{f_edge_solution} up to linear order in $t$ and $\mathcal{O}(\sqrt{\kappa})$,
while the black solid line corresponds to the solution \eqref{f_edge_solution} up to quadratic order in $t$ and 
$\mathcal{O}(\kappa^{\frac{3}{2}}\log \kappa)$.
It  can be observed that the former fits the numerical result well near the edge, 
while the latter provides an equally good fit, but over a wider region.
}\label{fig:edge_solution}
\end{center}
\end{figure}

\subsubsection{Numerical check}

We will check the above solution by comparing it with a numerical calculation.
In the numerical calculation, we have used the Gauss-Legendre quadrature with 1000 sample points
to approximate the integral \eqref{fredholm eq rescale f} \cite{Ling:2006up}.
In Figure \ref{fig:bulk_solution} and Figure \ref{fig:edge_solution}, 
we present the eigenvalue distribution obtained by the perturbative calculation and the numerical one for $\kappa=0.02$.
The figures demonstrate excellent agreement between the two.


\subsection{Solution of \eqref{Loop eq leading}} \label{app:solution of the full equation}

We consider
\begin{align}
\frac{1-\mathsf{D}^{\frac{\kappa}{2n}}}{1+\mathsf{D}^{\frac{\kappa}{2n}}}(1-\mathsf{D}^{\kappa}) R(x+i0)
+ \frac{1-\mathsf{D}^{- \frac{\kappa}{2n}}}{1+\mathsf{D}^{- \frac{\kappa}{2n}}}(1-\mathsf{D}^{- \kappa})R(x-i0) 
&= \frac{\pi\kappa}{n}x \quad (x\in [-1,1]),
\label{Loop eq leading rescale}
\end{align}
which is obtained from \eqref{Loop eq leading} with \eqref{R and R0} and \eqref{kappa_xm}. Here, $n\in \bm{Z}_{\geq 1}$.
We write the solution as 
\begin{align}
R(z)=-R^{[2]}(z)+\xi R^{[0]}(z),
\label{R_from_R0_and_R2}
\end{align}
where the constant $\xi$, which depends on $\kappa$ and $n$, is determined by $f(\pm 1)=0$.
$R^{[r]}(z)$ ($r=0,2$) are the resolvents that have a discontinuity along the interval $[-1,1]$ and satisfy,
\begin{align}
\frac{1-\mathsf{D}^{\frac{\kappa}{2n}}}{1+\mathsf{D}^{\frac{\kappa}{2n}}}(1-\mathsf{D}^{\kappa}) R^{[0]}(x+i0)
+ \frac{1-\mathsf{D}^{- \frac{\kappa}{2n}}}{1+\mathsf{D}^{- \frac{\kappa}{2n}}}(1-\mathsf{D}^{- \kappa})R^{[0]}(x-i0) 
&= 0, \label{R0 eq} \\
\frac{1-\mathsf{D}^{\frac{\kappa}{2n}}}{1+\mathsf{D}^{\frac{\kappa}{2n}}}(1-\mathsf{D}^{\kappa}) R^{[2]}(x+i0)
+ \frac{1-\mathsf{D}^{- \frac{\kappa}{2n}}}{1+\mathsf{D}^{- \frac{\kappa}{2n}}}(1-\mathsf{D}^{- \kappa})R^{[2]}(x-i0) 
&= -\frac{\pi\kappa}{n}x. \label{R2 eq}
\end{align}
We decompose the resolvent into the regular part and the discontinuous part, $R^{[r]}(z)=R^{[r]}_{\mathrm{reg}}(z)+R^{[r]}_{\mathrm{disc}}(z)$  ($r=0,2$).
From \eqref{R0 eq} and \eqref{R2 eq}, the regular part is obtained as
\begin{align}
R^{[0]}_{\mathrm{reg}}(z)&=\frac{\pi}{\kappa}z, \label{R[0]_reg} \\
R^{[2]}_{\mathrm{reg}}(z)&=\frac{\pi}{\kappa}\Big(\frac{1}{3}z^3-\frac{1}{2}z\Big). \label{R[2]_reg}
\end{align}
The discontinuous part should then satisfy
\begin{align}
\frac{1-\mathsf{D}^{\frac{\kappa}{2n}}}{1+\mathsf{D}^{\frac{\kappa}{2n}}}(1-\mathsf{D}^{\kappa})R^{[r]}_{\mathrm{disc}}(x+i0)
+\frac{1-\mathsf{D}^{- \frac{\kappa}{2n}}}{1+\mathsf{D}^{- \frac{\kappa}{2n}}}(1-\mathsf{D}^{- \kappa})R^{[r]}_{\mathrm{disc}}(x-i0)&=0,
\label{Loop eq leading rescale disc r} 
\end{align}
and 
\begin{align}
R^{[r]}_{\mathrm{disc}}(z)\to -R^{[r]}_{\mathrm{reg}}(z) + \mathcal{O}\big(\tfrac{1}{z}\big) \quad \mathrm{in} \; z\to \infty.
\label{R_asymptotic}
\end{align}
We denote by $f^{[r]}(z)$ the eigenvalue distribution corresponding to $R^{[r]}(z)$.
From $f(\pm 1)=0$, $\xi$ can be expressed as 
\begin{align}
\xi=\frac{f^{[2]}(1)}{f^{[0]}(1)}.
\label{xi_coeff}
\end{align}

\subsubsection{Solving the discontinuous part of $R^{[r]}(z)$}

We obtain the perturbative expansion for $R^{[r]}_{\mathrm{disc}}(z)$ in each of the bulk regime and the edge regime,
using the same method as in Appendix \ref{solution of fredholm eq}.

In the bulk regime, we put the ansatz for $r=0$ and $r=2$ of the form
\begin{align}
R_{\text{disc,b}}^{[0]}(x)
&=-\frac{\pi}{\kappa}(x^2-1)^{\frac{1}{2}}
+\sum_{l,m=0}^{\infty}\sum_{k=0}^{l+m+1}c^{[0]}_{lmk}\kappa^{l+m}\frac{x^{p_0(k)}}{(x^2-1)^{l+\frac{1}{2}}}
\log^k\Big(\frac{x-1}{x+1}\Big),\nonumber\\
R_{\text{disc,b}}^{[2]}(x)
&=-\frac{\pi}{3 \kappa}(x^2-1)^{\frac{3}{2}}
+\sum_{l,m=0}^{\infty}\sum_{k=0}^{l+m+1}c^{[2]}_{lmk}\kappa^{l+m}\frac{x^{p_0(k)}}{(x^2-1)^{l-\frac{1}{2}}}
\log^k\Big(\frac{x-1}{x+1}\Big).
\label{expansion_bulk_0_and_2}
\end{align}
The coefficients $c^{[r]}_{lmk}$ are polynomials of $\log(\kappa)$, dependent on $n$, and are determined by matching with the edge regime expansion.
Note that for \eqref{R_asymptotic} to hold, we must have $c^{[2]}_{0m0}=2c^{[2]}_{0m1}$.

In the edge regime, we consider $R_{\text{disc,e}}^{[r]}(t)$ and its Laplace transform $\hat{R}_{\text{disc,e}}^{[r]}(s)$, 
defined in \eqref{R0_edge_disc} and \eqref{Laplace_transform}, respectively.
From \eqref{Loop eq leading rescale disc r}, the discontinuity of $R_{\text{disc,e}}^{[r]}(t)$ ($t<0$) is given by
\begin{align}
\frac{1-\mathsf{D}^{\frac{1}{n}}}{1+\mathsf{D}^{\frac{1}{n}}}(1-\mathsf{D}^{2})R^{[r]}_{\mathrm{disc,e}}(t+i0)
+\frac{1-\mathsf{D}^{- \frac{1}{n}}}{1+\mathsf{D}^{- \frac{1}{n}}}(1-\mathsf{D}^{-2})R^{[r]}_{\mathrm{disc,e}}(t-i0)&=0.
\label{Loop eq leading rescale disc r edge} 
\end{align}
One finds from the expansion for $|t|\ll \frac{1}{n}$ that $R^{[r]}_{\mathrm{disc,e}}(t+i0)+R^{[r]}_{\mathrm{disc,e}}(t-i0)$ is analytic at $t=0$. 
This implies that $R^{[r]}_{\mathrm{disc,e}}(t)$ for $|t|\ll \frac{1}{n}$ is given by a power series in half-integer powers of $t$.
The Laplace transform $\hat{R}_{\text{disc,e}}^{[r]}(s)$ has a discontinuity along the negative real axis and satisfies
\begin{align}
\frac{1-e^{-i\frac{s}{n}}}{1+e^{-i\frac{s}{n}}}(1-e^{-2is})\hat{R}_{\text{disc,e}}^{[r]}(s-i0)
+\frac{1-e^{i\frac{s}{n}}}{1+e^{i\frac{s}{n}}}(1-e^{2is})\hat{R}_{\text{disc,e}}^{[r]}(s+i0)=0,
\label{eq_R0_edge_s_disc}
\end{align}
for $s<0$. 
The analytic form of $\hat{R}_{\text{disc,e}}^{[r]}(s)$ is determined order by order in the $\kappa$-expansion by the following properties:
It is analytic everywhere except on the negative real axis, where it satisfies \eqref{eq_R0_edge_s_disc}.
For small $s$, in order to ensure the matching with the bulk regime as discussed in Appendix \ref{Solving the discontinuous part of R[r](z)},
it should have an expansion in half-integer powers of $s$ greater than or equal to $-\frac{3}{2}-l$ for terms of $\mathcal{O}(\kappa^{l})$.
For large $s$, it admits a power series expansion in half-integer powers of $s$, 
as follows from the Laplace transform of the small $t$ expansion of $R^{[r]}_{\mathrm{disc,e}}(t)$ in half-integer powers of $t$,
together with the transform pair $s^{-a-\frac{1}{2}}\leftrightarrow \Gamma(-a+\tfrac{1}{2}) t^{a-\frac{1}{2}}$.

Hence, we adopt the following ansatz for $\hat{R}_{\text{disc,e}}^{[r]}(s)$:
\begin{align}
\hat{R}_{\text{disc,e}}^{[r]}(s)=
\frac{1}{\sqrt{\kappa}}
\Phi(s)
Q^{[r]}(s),
\label{expansion_edge_0}
\end{align}
where 
\begin{align}
\Phi(s)&\equiv \frac{1}{s^{\frac{3}{2}}}\exp\left[\frac{s}{\pi}\log\left(\frac{\pi e}{s}\right)\right]\Gamma\left(\frac{s}{\pi}+1\right)
\frac{\sqrt{\pi}\Gamma\left(\frac{s}{2n\pi}+1\right)}{\Gamma\left(\frac{s}{2n\pi}+\frac{1}{2}\right)}, \label{Phi(s)} \\
Q^{[r]}(s)&\equiv \sum_{l,m=0}^{\infty}Q^{[r]}_{lm}\frac{\kappa^{l+m}}{s^l}. \label{Qr(s)}
\end{align}
The coefficients $Q^{[r]}_{lm}$ are polynomials of $\log(\kappa)$ and dependent on $n$.
Note that from \eqref{expansion_edge_0}, $R^{[r]}_{\mathrm{disc,e}}(t)$ for $|t|\ll \frac{1}{n}$ is given by
\begin{align}
R^{[r]}_{\mathrm{disc,e}}(t)|_{|t|\ll \frac{1}{n}} \simeq \pi \sum_{a=0}^{\infty} (-1)^{a}f^{[r]}_at^{a-\frac{1}{2}},
\label{small t expansion of edgeR} 
\end{align}
where $f^{[r]}_a$ are the expansion coefficients.

The coefficients $Q^{[r]}_{lm}$ and $c^{[r]}_{lmk}$ can be fixed simultaneously by matching 
 the bulk expansion \eqref{expansion_bulk_0_and_2} and the edge expansion \eqref{expansion_edge_0}.
The coefficients for $r=0$, $Q^{[0]}_{lm}$ and $c^{[0]}_{lmk}$,  are determined by
\begin{align}
& Q^{[0]}_{00} = \frac{\sqrt{\pi}}{2}, \quad
 Q^{[0]}_{01} = - \frac{L}{8 \sqrt{\pi}} + \frac{C_{1}}{4\sqrt{\pi}n} ,\quad
 Q^{[0]}_{02} = - \frac{L^{2}}{64 \pi^{\frac{3}{2}}} - \frac{L}{16 \pi^{\frac{3}{2}}} 
                        + \frac{C_{1} L}{16 \pi^{\frac{3}{2}} n} + \frac{C_{1}}{8 \pi^{\frac{3}{2}} n} - \frac{C_{1}^{2}}{16\pi^{\frac{3}{2}} n^{2}}, \quad
 Q^{[0]}_{10} = - \frac{3\sqrt{\pi}}{32}, \nonumber\\
& Q^{[0]}_{11} =  - \frac{3 L}{128 \sqrt{\pi}} - \frac{5}{128 \sqrt{\pi}} + \frac{3 C_{1}}{64\sqrt{\pi} n}, \quad
 Q^{[0]}_{20} = - \frac{15\sqrt{\pi}}{1024}, \quad
 \nonumber \\[3mm]
& c^{[0]}_{000} = - \frac{L}{2} + \frac{1}{2} + \frac{C_{1}}{n}, \quad
 c^{[0]}_{001} = \frac{1}{2}, \quad
 c^{[0]}_{010} = \frac{L^{2}}{8 \pi} - \frac{1}{4 \pi} - \frac{C_{1} L}{2\pi n} + \frac{C_{1}^{2}}{2\pi n^{2}}, \quad
 c^{[0]}_{011} = 0, \nonumber\\
& c^{[0]}_{012} = 0, \quad
 c^{[0]}_{100} = \frac{L^{2}}{8 \pi} + \frac{L}{4 \pi} - \frac{3}{8 \pi} + \frac{\pi}{12} - \frac{C_{1} L}{2\pi n} - \frac{C_{1}}{2\pi n} + \frac{C_{2}}{n^{2}}, \quad
 c^{[0]}_{101} = - \frac{L}{4 \pi} - \frac{1}{4 \pi} + \frac{C_{1}}{2\pi n}, \quad
 c^{[0]}_{102} = \frac{1}{8 \pi}, 
\label{coeff r=0}
\end{align}
where $L\equiv \log\left(\frac{\kappa}{16\pi}\right)$. 
$C_1$ and $C_2$ are numerical constants defined by
\begin{align}
C_1&\equiv \frac{\log 2}{2}, 
\quad
C_2\equiv -\frac{\pi}{96}+\frac{(\log 2)^2}{8\pi}.
\end{align}
The coefficients for $r=2$, $Q^{[2]}_{lm}$ and $c^{[2]}_{lmk}$,  are obtained as
\begin{align}
& Q^{[2]}_{00} = 0,\quad
 Q^{[2]}_{01} = - \frac{L}{4 \sqrt{\pi}} - \frac{3}{4 \sqrt{\pi}} + \frac{C_{1}}{2\sqrt{\pi}n}, \quad
 Q^{[2]}_{02} = \frac{L^{2}}{8 \pi^{\frac{3}{2}}} + \frac{7 L}{16 \pi^{\frac{3}{2}}} - \frac{\sqrt{\pi}}{12} - \frac{3}{8 \pi^{\frac{3}{2}}} 
 - \frac{C_{1} L}{2\pi^{\frac{3}{2}} n} - \frac{7 C_{1}}{8 \pi^{\frac{3}{2}} n} 
 + \frac{C_{1}^{2}}{\pi^{\frac{3}{2}} n^{2}}  - \frac{C_{2}}{\sqrt{\pi} n^{2}}, \nonumber\\
& Q^{[2]}_{10} = -\frac{\sqrt{\pi}}{4}, \quad
 Q^{[2]}_{11} = \frac{15 L}{64 \sqrt{\pi}} + \frac{25}{64 \sqrt{\pi}} - \frac{15 C_{1}}{32 \sqrt{\pi} n}, \quad
 Q^{[2]}_{20} = \frac{15\sqrt{\pi}}{64}, \quad 
 \nonumber\\[3mm]
& c^{[2]}_{000} = 1, \quad
 c^{[2]}_{001} = \frac{1}{2}, \quad
 c^{[2]}_{010} = 0, \quad
 c^{[2]}_{011} = 0, \quad
 c^{[2]}_{012} = - \frac{1}{4\pi}, \nonumber\\
& c^{[2]}_{100} = \frac{L^{2}}{8 \pi} + \frac{3 L}{4 \pi} - \frac{7}{8 \pi} - \frac{\pi}{12}
 - \frac{C_{1} L}{2\pi n} - \frac{3 C_{1}}{2\pi n} 
 + \frac{C_{1}^{2}}{\pi n^{2}} - \frac{C_{2}}{n^{2}},\quad
 c^{[2]}_{101} = -\frac{1}{\pi}, \quad
 c^{[2]}_{102} = -\frac{1}{8\pi}.
\label{coeff r=2}
\end{align}
Note that in $n\to \infty$, $Q^{[r]}_{lm}$ and $c^{[r]}_{lmk}$ reduce to $\bar{Q}^{[r]}_{lm}$ and $\bar{c}^{[r]}_{lmk}$.

\subsubsection{Edge behavior of $f^{[r]}(x)$}

One finds from \eqref{small t expansion of edgeR} that 
the eigenvalue distribution in the edge regime, $f^{[r]}_{\text{e}}(t)=f^{[r]}(1+\frac{\kappa t}{2})$ with $\kappa\to 0$, takes the form for $|t|\ll \frac{1}{n}$ as
\begin{align}
f^{[r]}_{\text{e}}(t)\big|_{|t|\ll \frac{1}{n}}
\simeq \sum_{a=0}^{\infty} f_a^{[r]} |t|^{a-\frac{1}{2}}. \label{expansion_fr_edge}
\end{align}

We obtain $f_a^{[r]}$ in terms of $Q_{lm}^{[r]}$ by performing an expansion of the edge-regime ansatz \eqref{expansion_edge_0} for $s\gg n$. 
We denote by $g_k$ the expansion coefficients of \eqref{Phi(s)} in this regime,
\begin{align}
\Phi(s)\bigr|_{s\gg n}=\sum_{k=0}^\infty \frac{g_k}{s^{k+\frac{1}{2}}}
 \label{expansion_gamma}
\end{align}
The coefficients $g_k\; (k=0,1,2,\cdots)$ are given by
\begin{align}
g_0=\frac{1}{\sqrt{n}},\;\; 
g_1=\frac{\pi \sqrt{n}}{4} + \frac{\pi}{12 \sqrt{n}}, \;\;
g_2=\frac{\pi^{2} n^{\frac{3}{2}}}{32} + \frac{\pi^{2} \sqrt{n}}{48} + \frac{\pi^{2}}{288 \sqrt{n}},\;\; 
\cdots.
\label{gk}
\end{align}
We also define
\begin{align}
Q^{[r]}_{(a)}
&\equiv \sum_{l=0}^{a}\sum_{m=0}^{\infty} g_{a-l}\kappa^{l+m} Q^{[r]}_{lm} 
=g_aQ^{[r]}_{00}+\left(g_aQ^{[r]}_{01} + g_{a-1} Q^{[r]}_{10}\right)\kappa +\cdots,
\label{tildeQn}
\end{align}
and write the expansion of the edge-regime ansatz \eqref{expansion_edge_0} for $s\gg n$ as
\begin{align}
\hat{R}_{\text{disc,e}}^{[r]}(s)\Bigr|_{s\gg n}
\simeq \frac{1}{\sqrt{\kappa}}\sum_{a=0}^\infty \frac{Q^{[r]}_{(a)}}{s^{a+\frac{1}{2}}}.
\label{Rs_tildeQ}
\end{align}
By performing the (analytically continued) Laplace transform
and comparing with \eqref{small t expansion of edgeR}, one finds
\begin{align}
f_a^{[r]}
=\frac{(-1)^a}{\pi\sqrt{\kappa}}\Gamma\left(-a+\tfrac{1}{2}\right)Q^{[r]}_{(a)}.
\label{f_a}
\end{align}

\subsubsection{Result for $f(x)$ in the strong coupling expansion}

From \eqref{coeff r=0}--\eqref{f_a}, one can evaluate $\xi$ in \eqref{xi_coeff} as
\begin{align}
\xi
&=\kappa \left(- \frac{\log\left(\frac{\kappa}{16\pi}\right)}{2 \pi} - \frac{3}{2 \pi} + \frac{C_{1}}{\pi n}\right) \nonumber\\
&\quad + \kappa^{2} \left(
\frac{\log\left(\frac{\kappa}{16\pi}\right)^{2}}{8 \pi^{2}} + \frac{\log\left(\frac{\kappa}{16\pi}\right)}{2 \pi^{2}} - \frac{1}{6} - \frac{3}{4 \pi^{2}}
 - \frac{C_{1} \log\left(\frac{\kappa}{16\pi}\right)}{2\pi^{2} n} - \frac{C_{1}}{\pi^{2} n}
+\frac{3C_{1}^{2}}{2\pi^2 n^{2}} - \frac{2 C_{2}}{\pi n^{2}} 
\right) 
+ \mathcal{O}\left(\kappa^{3}(\log\kappa)^3\right).
\label{gamma_solution}
\end{align}
\subsubsection*{$\bullet$ resolvent $R(z)$}
The regular part of $R(z)$ is obtained from \eqref{R_from_R0_and_R2}, \eqref{R[0]_reg} and \eqref{R[2]_reg} as
\begin{align}
R_{\mathrm{reg}}(z)
= \frac{\pi}{\kappa}\left(
-\frac{1}{3}z^3+\left(\frac{1}{2}+\xi\right)z
\right).
\end{align}
The discontinuous part in the bulk regime is given by
\begin{align}
R_{\text{disc,b}}(z)
=
\frac{\pi}{3\kappa} (z^2-1)^{\frac{3}{2}} 
+\left(\frac{\log\left(\frac{\kappa}{16\pi}\right)}{2} + \frac{1}{2} - \frac{C_{1}}{n}\right)\sqrt{z^2-1}
-\frac{1}{2}z\sqrt{z^2-1}\ln\left(\frac{z-1}{z+1}
\right)
+\mathcal{O}(\kappa(\log\kappa)^2).
\end{align}
The discontinuous part in the edge regime, $R_{\text{disc,e}}(t)=R_{\text{disc}}(1+\frac{\kappa t}{2})$, is given by
\begin{align}
&R_{\text{disc,e}}(t)=\int_0^{\infty} ds\, e^{-ts}\hat{R}_{\text{disc,e}}(s),
\end{align}
with 
\begin{align}
\hat{R}_{\text{disc,e}}(s)
&= 
\frac{1}{\sqrt{\kappa}}\left[
 \frac{\sqrt{\pi}}{4}\frac{\kappa}{s} 
+ \Bigl(- \frac{3 \log\left(\frac{\kappa}{16\pi}\right)}{16 \sqrt{\pi}} - \frac{1}{4 \sqrt{\pi}} +\frac{3C_1}{8\sqrt{\pi}n}\Bigr)\frac{\kappa^2 }{ s } 
- \frac{15 \sqrt{\pi}}{64}\frac{\kappa^2 }{ s^2 }
\right]
\Phi(s)
\end{align}
For $|t|\ll \frac{1}{n}$, it reduces to
\begin{align}
R_{\text{disc,e}}(t)|_{|t|\ll \frac{1}{n}} &= \pi \left(-f_1 t^{\frac{1}{2}}+f_2t^{\frac{3}{2}}+\cdots\right),
\end{align}
where 
\begin{align}
f_1&=
\frac{1}{2\sqrt{n}}\sqrt{\kappa}
-\frac{1}{\pi} \left( 
 \frac{3 \log\left(\frac{\kappa}{16\pi}\right)}{8 \sqrt{n}} + \frac{1}{2\sqrt{n}} - \frac{3 C_{1}}{4n^{\frac{3}{2}}}  
\right)\kappa^{\frac{3}{2}} + \mathcal{O}\left(\kappa^\frac{5}{2}(\log\kappa)^2\right), \nonumber\\
f_2&=
\pi \left( \frac{\sqrt{n}}{12} + \frac{1}{36\sqrt{n}} \right)\sqrt{\kappa}
- \left(
\frac{\sqrt{n} \log\left(\frac{\kappa}{16\pi}\right)}{16} + \frac{\sqrt{n}}{12}
 + \frac{\log\left(\frac{\kappa}{16\pi}\right)}{48\sqrt{n}} + \frac{49}{144\sqrt{n}} - \frac{C_{1}}{8\sqrt{n}}  
 - \frac{C_1}{24n^{\frac{3}{2}}} 
\right)\kappa^{\frac{3}{2}} \nonumber\\
&\quad +\mathcal{O}\left(\kappa^\frac{5}{2}(\log\kappa)^2\right).
\label{barf1_and_barf2}
\end{align}

\subsubsection*{$\bullet$ eigenvalue distribution $f(x)$}

The eigenvalue distribution in the bulk regime is given by
\begin{align}
f_{\mathrm{b}}(x)
&=\frac{1}{3 \kappa}(1-x^2)^{\frac{3}{2}}
-\frac{1}{\pi}\left(\frac{\log\left(\frac{\kappa}{16\pi}\right)}{2} + \frac{1}{2} - \frac{C_{1}}{n}\right)\sqrt{1-x^2}
+\frac{1}{2\pi}x\sqrt{1-x^2}\ln\left(\frac{1-x}{1+x}\right)
+\mathcal{O}\big(\kappa(\log\kappa)^2\big). \label{f_bulk_solution} 
\end{align}
The eigenvalue distribution in the edge regime, $f_{\mathrm{e}}(t)=f(1+\frac{\kappa t}{2})$, is given by
\begin{align}
f_{\mathrm{e}}(t)|_{t\ll \frac{1}{n}}&=  f_1|t|^{\frac{1}{2}}+f_2|t|^{\frac{3}{2}}+\cdots, \label{f_edge_solution} 
\end{align}
where $f_1$ and $f_2$ are provided in \eqref{barf1_and_barf2}.

\subsubsection*{$\bullet$ the extent of the eigenvalues $x_m$}

Finally, we present $x_m$ as a function of $\lambda$.
From the large $z$ expansion of the resolvent in the bulk regime, we obtain $T_0(\kappa,n)$ in \eqref{Ta} as
\begin{align}
T_0(\kappa,n)=\frac{\pi}{8\kappa}-\frac{\log\left(\frac{\kappa}{16\pi}\right)}{4}-\frac{5}{12}+\frac{C_1}{2n}+\mathcal{O}(\kappa(\log\kappa)^2).
\label{T0_solution}
\end{align}
By using \eqref{lambda_xm} and \eqref{T0_solution}, 
$\lambda$ can be expressed as a function of $\kappa$.
Then, with \eqref{kappa_xm}, $x_m$ can be obtained as a function of $\lambda$,
\begin{align}
x_m=(8\lambda)^{\frac{1}{4}}+\frac{n}{\pi}\left(-\log\frac{8\pi(8\lambda)^{\frac{1}{4}}}{n}+\frac{5}{3}\right)-\frac{2C_1}{\pi}
+\mathcal{O}\left(\lambda^{-\frac{1}{4}}(\log\lambda)^2\right).
\end{align}


\section{Relation between $\hat{\bm{\mathsf{D}}}_{n_1,n_2} R(x)$ and $\hat{\bm{\mathsf{D}}}_{n_1} R(x)$} 
\label{app:derivation_S}

In this appendix, we prove \eqref{hatR1 to hatR2}, which holds for $n_1>n_2$.
For $z_{\pm}\in \mathbb{C}$, $\operatorname{Im}z_{\pm}\gtrless 0$,  one can show the equality
\begin{align}
\frac{1}{2\pi i}\int_{-\infty}^{\infty}dx\frac{1}{z_{\pm}-x}\hat{\bm{\mathsf{D}}}_{n_1} R(x)
&=\mp \sum_{J=0}^{n_1-1}\left\{R(z_{\pm}\pm 2Ji)+R(z_{\pm}\pm (2J+2)i)-2R(z_{\pm}\pm (2J+1)i)\right\} \nonumber\\
&=\mp \frac{1-\mathsf{D}^{\pm 1}}{1+\mathsf{D}^{\pm 1}}(1-\mathsf{D}^{\pm 2n_1})R(z_{\pm}\pm i\epsilon).
\label{useful eq 0}
\end{align}
Suppose $y\in \mathbb{R}, \: 0<t<2n_1$. 
Summing up \eqref{useful eq 0} with $z_+=y+i(t+2kn_1)  \; (k=0,1,\cdots)$, we obtain
\begin{align}
\frac{1}{2\pi i}\int_{-\infty}^{\infty}dx\sum_{k=0}^{\infty} \frac{1}{y +  i(t+2kn_1) - x}\hat{\bm{\mathsf{D}}}_{n_1} R(x)
&=-\frac{1-\mathsf{D}}{1+\mathsf{D}}R(y+it). \label{pos sum}
\end{align}
Likewise, the sum of \eqref{useful eq 0} with $z_-=y+i(t-2kn_1)  \; (k=1,2,\cdots)$ gives
\begin{align}
\frac{1}{2\pi i}\int_{-\infty}^{\infty}dx\sum_{k=1}^{\infty} \frac{1}{y +  i(t-2kn_1) - x}\hat{\bm{\mathsf{D}}}_{n_1} R(x)
&=\frac{1-\mathsf{D}^{-1}}{1+\mathsf{D}^{-1}}R(y-i(2n_1-t)). \label{neg sum}
\end{align}
Adding \eqref{pos sum} and \eqref{neg sum} results in
\begin{align}
\frac{1}{2\pi i}\int_{-\infty}^{\infty}dx\sum_{k=-\infty}^{\infty} \frac{1}{y +  i(t+2kn_1) - x} \hat{\bm{\mathsf{D}}}_{n_1} R(x)
=-\frac{1-\mathsf{D}}{1+\mathsf{D}}R(y+it)+\frac{1-\mathsf{D}^{-1}}{1+\mathsf{D}^{-1}}R(y+i(t-2n_1)).
\label{useful eq 1}
\end{align}
From \eqref{useful eq 1} with $t=n_1-n_2$ and $t=n_1+n_2$, where $n_1>n_2$, we can show
\begin{align}
\hat{\bm{\mathsf{D}}}_{n_1,n_2} R(y)
=\frac{1}{2n_1}\int_{-\infty}^{\infty} dx \frac{\sin\frac{\pi n_2}{n_1}}{\cosh\frac{\pi(y-x)}{n_1}+\cos\frac{\pi n_2}{n_1}}\hat{\bm{\mathsf{D}}}_{n_1} R(x),
\end{align}
where we used
\begin{align}
\sum_{k=-\infty}^{\infty}\frac{1}{y+i(n_1+n_2)-x+2in_1k}-\frac{1}{y+i(n_1-n_2)-x+2in_1k}
=\frac{i\pi}{n_1}\frac{\sin\frac{\pi n_2}{n_1}}{\cosh\frac{\pi(y-x)}{n_1}+\cos\frac{\pi n_2}{n_1}}.
\end{align}

Note that for $n_1<n_2$, the relationship between $\hat{\bm{\mathsf{D}}}_{n_1,n_2} R(x)$ and $\hat{\bm{\mathsf{D}}}_{n_1} R(x)$ takes a different form: 
\begin{align}
\hat{\bm{\mathsf{D}}}_{n_1,n_2} R(y)
&=\frac{1-\mathsf{D}}{1+\mathsf{D}}(\mathsf{D}^{n_2-n_1}-\mathsf{D}^{n_2+n_1})R(y+i0)
+\frac{1-\mathsf{D}^{-1}}{1+\mathsf{D}^{-1}}(\mathsf{D}^{-n_2+n_1}-\mathsf{D}^{-n_2-n_1})R(y-i0) \nonumber\\
&=\frac{1}{\pi}\int_{-\infty}^{\infty} dx  \frac{n_2-n_1}{(y-x)^2+(n_2-n_1)^2} \hat{\bm{\mathsf{D}}}_{n_1} R(x).
\label{S n1<n2}
\end{align}

\bibliographystyle{utphys}
\bibliography{dsl_in_pwmm} 

\end{document}